\RequirePackage{lineno}
\documentclass[onecolumn,preprintnumbers,amsmath,amssymb,nofootinbib]{revtex4}
\usepackage[utf8]{inputenc}

\usepackage{xfrac} 
\usepackage{amsmath}
\usepackage{float}
\usepackage{dsfont} 
\usepackage{physics} 
\usepackage{braket}
\usepackage{gensymb} 
\usepackage{tensor} 
\usepackage{xcolor,soul}
\sethlcolor{yellow}

\usepackage{epsfig}
\usepackage{graphicx}
\usepackage{dcolumn}
\usepackage{bm}
\usepackage{amsmath}
\usepackage{amsthm}
\usepackage{amsfonts}
\usepackage{color} 
\usepackage{caption}
\usepackage{subcaption}
\usepackage{siunitx,booktabs}
\usepackage{hyperref}
\usepackage{epstopdf}

\allowdisplaybreaks

\begin{document}
\title{Quantum key distribution with flawed and leaky sources}
\author{Margarida Pereira$^{1}$}
\email{mpereira@com.uvigo.es}
\author{Marcos Curty$^{1}$, Kiyoshi Tamaki$^{2}$}
\affiliation{$^{1}$Escuela de Ingenier$\acute{\textit{\i}}$a de Telecomunicaci$\acute{o}$n, Department of Signal Theory and Communications, University of Vigo, Vigo E-36310, Spain \\
$^{2}$Graduate School of Science and Engineering for Research,University of Toyama, Gofuku 3190, Toyama 930-8555, Japan}
\date{\today}

\begin{abstract}
In theory, quantum key distribution (QKD) allows secure communications between two parties based on physical laws. However, most of the security proofs of QKD today make unrealistic assumptions and neglect many relevant device imperfections. As a result, they cannot guarantee the security of the practical implementations. Recently, the loss-tolerant protocol (K. Tamaki et al, Phys. Rev. A, 90, 052314, 2014) was proposed to make QKD robust against state preparation flaws. This protocol relies on the emission of qubit systems which, unfortunately, is difficult to achieve in practice. In this work, we remove such qubit assumption and generalise the loss-tolerant protocol to accommodate multiple optical modes in the emitted signals. These multiple optical modes could arise, for example, from Trojan horse attacks and/or device imperfections. Our security proof determines some dominant device parameter regimes needed for achieving secure communication, and therefore it can serve as a guideline to characterise QKD transmitters. Furthermore, we compare our approach with that of Lo and Preskill (H.-K. Lo et al, Quantum Inf. Comput., 7, 431-458, 2007) and identify which method provides the highest secret key generation rate as a function of the device imperfections. Our work constitutes an important step towards the best practical and secure implementation for QKD. 
\end{abstract}

\maketitle

\section{Introduction}
\label{sec:intro}
Quantum key distribution (QKD)  \cite{gisin,scarani,lo} enables two distant parties, Alice and Bob, to share a common secret key that can be used to encrypt and decrypt messages. In theory, QKD can offer information-theoretic security based on the laws of physics. In practise, however, it does not because typical security proofs of QKD require assumptions that are not actually met by the practical implementations, as they usually ignore many experimental device imperfections. This discrepancy between the theory and the practice of QKD has been evidenced by many quantum hacking attacks, especially by those that exploit flaws in the detectors of QKD systems \cite{lydersen,gerhardt}. Fortunately, the proposal of measurement-device-independent QKD (MDI-QKD) \cite{lo2} can solve all security loopholes in the measurement unit, and therefore Eve cannot take advantage of detector side-channels to learn information about the key. Furthermore, MDI-QKD can be implemented experimentally using standard optical components \cite{rubenok,silva,liu,tang,yin,roberts}. Therefore, to guarantee implementation security we now need to focus on how to secure the source in QKD.   

Ideally, the sending devices are single-photon sources and the encoding of the light pulses is executed perfectly, without any state preparation flaw (SPF). However, none of these two conditions are met experimentally since all devices have inherent deficiencies. The decoy-state method \cite{hwang,lo3,wang2} was proposed to replace single-photon sources with coherent light sources. Also, by using the Gottesman-Lo-L{\"u}tkenhaus-Preskill (GLLP) security analysis \cite{gottesman} the problem with SPFs is fixed. The main drawback of this last approach is that the resulting secret key rate is poor and fragile against channel loss. This is because it assumes the worst case scenario in which Eve could enhance the signals' flaws through channel loss, which significantly decreases the performance of the QKD scheme. 

Recently, a protocol that is loss-tolerant to SPFs has been proposed \cite{tamaki} to address the limitation of the GLLP analysis. The loss-tolerant protocol employs only three states and takes into account modulation errors due to an imperfect phase modulator (PM). Remarkably, by using a different phase error estimation technique involving the use of the basis mismatched events, the secret key rate of the loss-tolerant protocol remains almost unchanged even if the SPFs increase. In fact, the maximum transmission distance for a QKD system in fiber so far has been recently achieved using this protocol \cite{boaron}, which shows that the loss-tolerant protocol is highly practical. Its main weakness, however, is the assumption that the single-photon signals sent by Alice are qubits, which is difficult to guarantee in practice. For instance, if Eve conducts a Trojan horse attack (THA)  \cite{gisin2,vakhitov,lucamarini,tamaki3,wang} against the source, this assumption can be violated. In a THA, Eve sends bright light into Alice's PM and obtains information about the encoding by measuring the back-reflected light that exits Alice's lab. Moreover, an optical mode of the light pulse emitted by Alice could be dependent on the value of the phase modulation, which means that a sent single-photon pulse might not be a qubit (we call this imperfection the non-qubit assumption). That is, Alice's setting choice information could be encoded in other degrees of freedom of the emitted light, and this spontaneous leakage of information results in a higher-dimensional sending state. \\

This work aims to reduce the big gap between the theory and the practice of QKD by generalising the loss-tolerant protocol such that it can include typical imperfections in the sending device. To be precise, and in contrast to~\cite{tamaki}, here we remove the qubit assumption and include the effect of side-channels by considering the mode dependency of the PM and THAs. Moreover, like in~\cite{tamaki}, we also include in the analysis SPFs in a single-mode qubit subspace. Therefore, our analysis covers dominant imperfections that a source device has, allowing the use of a much wider class of imperfect devices in a secure manner. Our generalised loss-tolerant protocol can be applied to any multi-mode scenario as long as the states of the emitted signals are independently and identically distributed (I.I.D.), namely, this proof does not consider correlations between the sending signals. However, we remark that recent results reported in \cite{mizutani} imply that our analysis could also accommodate correlations between the signals which are independent of Alice's setting choice. Furthermore, we emphasise that the basic idea is rather general, and can be applied to many other QKD protocols like, for instance, the six-state protocol \cite{bruss}, distributed-phase-reference protocols \cite{inoue,takesue,stucki} and MDI-QKD~\cite{lo2}. In simple terms, it is a formalism to estimate the phase error rate of a QKD protocol by evaluating the transmission rates of some virtual states with the help of the state structure of Alice's signals (see Eq. (\ref{eq:multi-mode}) below). We also emphasise that our method does not require a complete characterisation of the side-channels, which significantly simplifies the experiments for characterising the source. Using this formalism, we can quantify the device parameters required to ensure secure communications with flawed and leaky sources. 

Additionally, we investigate how Lo-Preskill's security analysis \cite{lo4} behaves in the presence of the same device's imperfections and, by using imperfectly characterised states, we compare it with our generalised loss-tolerant protocol. As a result, we determine which security proof provides a higher secret key rate as a function of the device parameters. These parameters are essential for experimentalists to produce and to calibrate the transmitting devices, and therefore our work can be used as a guideline for securing the source in the presence of multi-mode signals.

This paper is organised as follows. In section \ref{sec:description}, we describe the assumptions that we make in our security proof and introduce the QKD protocol considered. In section \ref{sec:estimation}, we present the security analysis for our generalised loss-tolerant protocol. Then, in section \ref{sec:simulation}, we consider a particular device model to perform the simulations for our analysis and in order to compare it with Lo-Preskill's analysis. Finally, we conclude our paper in section \ref{sec:conclusion} and provide additional explanations and calculations in the Appendixes.

\section{Description of the protocol}
\label{sec:description}
We shall assume, for simplicity, that Alice's lab has a single-photon source. However, we emphasise that our analysis can also be applied to the case where Alice emits phase-randomised weak coherent pulses. In this latter case, Alice can use the decoy-state method \cite{hwang,lo3,wang2} to estimate all the quantities corresponding to the single-photon pulses which are needed to apply our method. Below we focus on the case where Alice has at her disposal single-photon sources only because the study with phase-randomised weak coherent pulses, together with decoy states, results in an unnecessarily cumbersome analysis. Fig. 1 shows the QKD setup. Next, we describe the assumptions we make on Alice's and Bob's devices.

\subsection{Assumptions on Alice's device}
In this work, we consider the asymptotic scenario where Alice sends Bob an infinite number of pulses. Our formalism is valid for any source that emits pulses whose quantum state is of the form
\begin{equation}
\ket{\Phi_{j\beta}}_{BE} = a_{j\beta} \ket{\phi_{j\beta}}_{BE} + b_{j\beta} \ket{\phi_{j\beta}^\perp}_{BE},
\label{eq:multi-mode}
\end{equation}
with $|a_{j\beta}|^2 + |b_{j\beta}|^2 =1$, where $j \in \{0,1\}$ and $\beta \in \{Z,X\}$ are Alice's bit value and basis choices respectively. As in the loss-tolerant analysis introduced in \cite{tamaki}, we consider a three-state protocol where Alice selects $j\beta \in \{0Z,1Z,0X\}$. Furthermore, in Eq. (\ref{eq:multi-mode}), we assume that $\ket{\phi_{j\beta}}_{BE}$ is a pure state in a single-mode qubit space, where $BE$ stands for Bob's and Eve's systems due to a potential THA. For instance, $\ket{\phi_{j\beta}}_{BE}$ could be of the form $\ket{\phi_{j\beta}}_{BE} = \ket{\omega_{j\beta}}_{B}\otimes \ket{\epsilon}_E$, where Eve's system does not depend on $j\beta$ and $\ket{\omega_{j\beta}}_{B}$ is a qubit state. The state $\ket{\phi_{j\beta}^\perp}_{BE}$, on the other hand, corresponds to any state outside of the single mode qubit space, including the state of a side-channel, and it is in an Hilbert space orthogonal to $\ket{\phi_{j\beta}}_{BE}$. We note that the form of the pure state given by Eq.~(\ref{eq:multi-mode}) is the most general I.I.D. state. Indeed, this equation simply decomposes a state in a given Hilbert space into a direct sum of two states in different Hilbert spaces, which can always be done. One of these states is in a qubit space and the other one is in any complementary Hilbert space. That is, any pure state can be written in the form given by Eq.~(\ref{eq:multi-mode}). In addition, 
\begin{figure}[H]
	\centering
	\includegraphics[width=17cm]{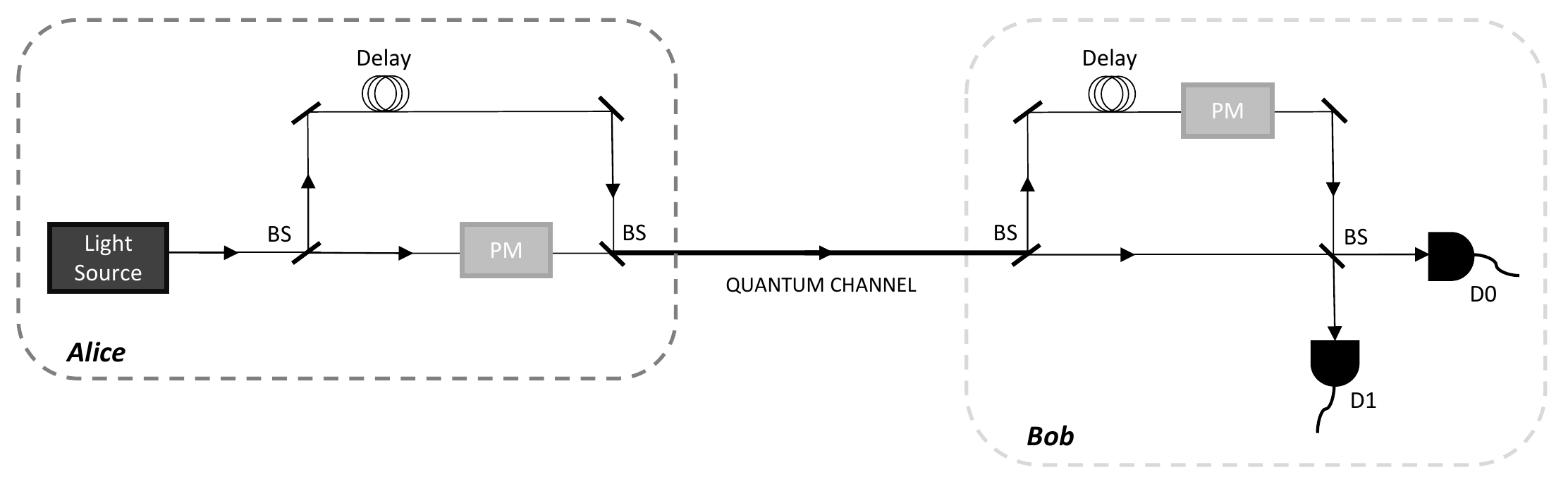} 
	\caption{Each single-photon pulse emitted by Alice's source goes through a 50:50 beamsplitter (BS) and is decomposed into the reference and the signal pulses. The reference pulse travels through the longer arm of Alice's Mach-Zehnder interferometer. To perform the encoding, she uses a PM that applies a phase shift to the signal pulse. The two pulses are recombined at the second 50:50 BS, sent through the quantum channel and then received in Bob's lab. On reception, they are split by a 50:50 BS and Bob applies a phase shift on the reference and signal pulses in the upper arm of his Mach-Zehnder interferometer. These pulses then interfere with the pulses that travelled through the shorter arm of the interferometer at the second 50:50 BS. Bob can then detect click events corresponding to photons choosing the shortest arm in Alice's interferometer and the longest one in Bob's, and to the opposite, by using two detectors, D0 and D1, which correspond to obtaining bit value 0 and 1 respectively. }
	\label{fig:protocol}
\end{figure} 
\noindent we further assume that, like in~\cite{tamaki}, the states $\ket{\phi_{j\beta}}_{BE}$ in Eq.~(\ref{eq:multi-mode}) form a triangle in the Bloch sphere, and we set their $Y$-components to be the same by choosing the $Y$-axis appropriately. This assumption is required to ensure that Alice is sending essentially three different states, rather than one or two states. Importantly, we note that  by introducing an ancilla system for Alice to purify the state, our formalism is also valid for a mixed state in a single-mode qubit space, as shown in Appendix \ref{app:security}.

The state structure in Eq. (\ref{eq:multi-mode}) means that the inner product ${\braket{\phi_{j\beta} | \phi_{j'\beta'}^\perp}}{_{BE}}$ for all $j,j',\beta$ and $\beta'$ is always zero. Also, depending on Alice's knowledge about the state given in Eq. (\ref{eq:multi-mode}) she might have to consider the worst-case scenario, {\it {\it i.e.}}, ${\braket{\phi_{j\beta}^\perp | \phi_{j'\beta'}^\perp}}{_{BE}} = 0$ for any combination of $(j,\beta)$ and $(j',\beta')$. This means that, complete information about $\ket{\phi_{j\beta}^\perp}$ is not required, which significantly simplifies the experiments for characterising the source. On the other hand, if Alice knows some structure of the side-channel she should fully exploit it and lower bound ${\braket{\phi_{j\beta}^\perp | \phi_{j'\beta'}^\perp}}{_{BE}}$. For example, if she knows that the side-channel is associated with the polarisation state of the single-mode qubit then the worst case scenario does not apply, {\it {\it i.e.}}, ${\braket{\phi_{j\beta}^\perp | \phi_{j'\beta'}^\perp}}{_{BE}} \neq 0$, since it is impossible for three states to be orthogonal to each other given the two-dimensionality of polarisation. This way, our formalism can readily take into account the available information.

We remark that, to apply the procedure introduced below we only need to determine the coefficients $a_{j\beta}$ and $b_{j\beta}$, and the qubit state, but it is not necessary to completely characterise the quantum information of the side-channel, $\ket{\phi_{j\beta}^\perp}_{BE}$. That is, our characterisation seems to be rather simple, and there is no need to perform further detailed characterisations. Nonetheless, the better Alice and Bob know the state given in Eq. (\ref{eq:multi-mode}), the better the resulting performance, as explained later in Section \ref{sec:simulation}. An experimental procedure to perform this estimation is out of the scope of this paper hence, we assume that these parameters are given. 

Furthermore, our work can accommodate any SPF in the single-mode qubit space, and one could also employ the techniques in \cite{nagamatsu,mizutani}. For example, we may select a case in which the states that Alice prepares can be expressed as 
\begin{linenomath*}
\begin{equation}
\frac{1}{\sqrt{2}}\left( \ket{1}_r \ket{v}_s + e^{\varphi_A + \delta \varphi_A /\pi} \ket{v}_r \ket{1}_s \right),
\label{eq:model}
\end{equation}\end{linenomath*}
where $\delta(\ge 0)$ is the deviation of the phase modulation from the intended value $\varphi_A$, and we define $\ket{1}_r \ket{v}_s = \ket{0_Y}$ and $ \ket{v}_r \ket{1}_s = \ket{1_Y}$, where $v$ stands for vacuum, $\ket{1}$ denotes a Fock state with one photon and the subscript $r$ $(s)$ corresponds to the reference (signal) pulse. In the case of the three-state protocol we have that $\varphi_A\in\{0, \pi, \pi/2\}$. Then, by using $\ket{0_Z} = (\ket{0_Y} + \ket{1_Y})/\sqrt{2}$ and $\ket{1_Z} = (- \ket{0_Y} + \ket{1_Y})/\sqrt{2}$, we obtain the following expressions for the three states in the single-mode qubit space:
\begin{linenomath*}
\begin{equation}
\begin{split}
&\ket{\omega_{0Z}}_B = \ket{0_Z}, \\
&\ket{\omega_{1Z}}_B = - \sin (\frac{\delta}{2})\ket{0_Z} +  \cos (\frac{\delta}{2}) \ket{1_Z},\\ 
&\ket{\omega_{0X}}_B = \cos (\frac{\pi}{4} + \frac{\delta}{4}) \ket{0_Z} +  \sin (\frac{\pi}{4} + \frac{\delta}{4}) \ket{1_Z}.
\end{split}
\label{eq:alice_states}
\end{equation}\end{linenomath*}

Therefore, our formalism can be used, for instance, when the information about Alice's choice of state is leaked and/or the optical mode depends on Alice's selection. This leakage from the source can occur spontaneously or with an active THA.

\subsection{Assumptions on Bob's device}
Bob receives the signal and reference pulses from Alice and measures them in a basis selected at random. More precisely, Bob's measurements are defined by the positive-operator valued measures (POVMs) $\{\hat{M}_{0\beta}, \hat{M}_{1\beta},\hat{M}_{f}\}$, where $\hat{M}_{0\beta}$ ($\hat{M}_{1\beta}$) with $\beta \in \{X,Z\}$ corresponds to obtaining the bit value 0 (1) when Bob chooses the basis $\beta$, and $\hat{M}_{f}$ corresponds to an inconclusive outcome. Importantly, $\hat{M}_{f}$ is assumed to be the same for the two bases. This means that the detection efficiencies are independent of Bob's measurement basis choice, which is required to prevent side-channel attacks exploiting channel loss \cite{lydersen,gerhardt}. Note that, this assumption is widely used in most security proofs, and one of the simplest ways to circumvent such detector side-channel attacks is to use MDI-QKD, to which our technique also applies (see Appendix \ref{app:mdi}).

\subsection{Actual protocol}

The three-state protocol makes an asymmetric choice of basis, namely, the $Z$ and the $X$ basis are selected with a priori different probabilities for both Alice and Bob. The events when both of them select the $Z$ basis are used for key generation. Also, all announcements between Alice and Bob are done via an authenticated public channel. Next, we describe the steps of the QKD protocol in detail. \\
 
\begin{enumerate}
\item \textbf{Initialisation:} Before running the protocol, Alice and Bob agree on a number $N_{fixed}$ of rounds, on the error correcting codes, and on a set of hash functions to perform privacy amplification. Steps 2-4 of the protocol are repeated $N$ times until the number of detected events $N$ becomes $N_{fixed}$.

\item \textbf{State preparation:} Alice generates the states given in Eq. (\ref{eq:multi-mode}). First, she selects the basis $\beta \in\{ Z, X\}$ for encoding the states with probabilities $P_{Z_A}$ and $P_{X_A} = 1- P_{Z_A}$ respectively. If the $Z$ basis is selected, she randomly chooses a bit value. Then, Alice prepares the signal and reference pulses following these specifications and sends the pulses to Bob via a quantum channel. Due to a potential THA, the sending states might contain Eve's system $E$ as well. 

\item \textbf{Measurement:} Bob measures each incoming signal using the basis $\beta \in \{Z,X\}$ which he selects with probabilities $P_{Z_B}$ and $P_{X_B} = 1 - P_{Z_B}$ respectively. 

\item \textbf{Detection announcement:} Bob checks whether the signal in Step 3 is detected or not. For each detected event, $N$ is increased by 1 unit\footnotemark[1]. If $N = N_{fixed}$, Bob announces the detected events, and they proceed to Step 5, otherwise they go back to Step 2.\footnotetext[1]{As proven in \cite{tamaki4}, one can employ this announcement and the setting independent termination condition as long as we use the Azuma's inequality \cite{azuma}. Moreover, with the Azuma's inequality, Alice and Bob can exchange the basis information for each round of the quantum communication \cite{tamaki4}, which we do not adopt here for simplicity of the discussions.}

\item \textbf{Basis announcement and sifting:} Alice and Bob announce their basis choices for the detected events. Also, they define bit strings associated with the basis matched and mismatched events.

\item \textbf{Parameter estimation:} Alice and Bob announce the bit strings $\vec{s}_{X,0Z}, ~\vec{s}_{X,1Z}$ and $\vec{s}_{X,0X}$ which correspond to the events when Alice sends one of the three possible states and Bob measures it in the $X$ basis. These bit strings are used to estimate the number of bits that need to be removed from the sifted key, which is composed by the $Z$ basis matched events, during privacy amplification. 

\item \textbf{Error correction and privacy amplification:} Alice and Bob randomly select an error correcting code from Step 1 to perform error correction on these sifted strings and then they exchange the syndrome information about the $Z$ basis matched events. Then, based on the result of the parameter estimation in Step 6, they perform privacy amplification on the corrected sifted keys. At the end of this step, Alice and Bob obtain the key strings $\vec{k}_{Z_A}$ and $\vec{k}_{Z_B}$, respectively. 

\end{enumerate} 

\vspace{3pt}
\section{Security Analysis}
\label{sec:estimation}
In order to prove the information-theoretic security of our protocol we use the complementary scenario introduced by Koashi \cite{koashi,koashi2}. For this, we first need to create an equivalent virtual protocol (see Appendix \ref{app:security}) concerning an observable conjugate to the key. The classical and quantum information available to Eve in the actual and virtual protocols are the same and therefore she cannot distinguish and behave differently between them. Hence, by proving the security in the virtual protocol we ensure the security of the actual protocol. Additionally, from the virtual protocol we can determine the phase error rate, which quantifies the amount of information that is leaked to Eve and has to be removed in the privacy amplification step. In this section, we show how this last quantity is estimated by generalising the loss-tolerant method.

\subsection{Secret key rate}
As explained before, for simplicity we assume the asymptotic scenario where Alice sends Bob an infinite number of pulses. The asymptotic key rate for the single-photon signals can be expressed as 
\begin{linenomath*}
\begin{equation}
R \ge Y_Z [1 - h(e_X) - fh(e_Z)],
\label{eq:keyrate}
\end{equation}\end{linenomath*}
where $Y_Z$ is the yield of the single photons in the $Z$ basis, {\it {\it i.e.}} the joint probability of Alice emitting a single-photon in the $Z$ basis and Bob detecting it with a measurement also in the $Z$ basis. The function $h(x) = -x\log_2 (x) - (1-x) \log_2 (1-x)$ is the binary entropy function and $f$ is the error correction efficiency. The term $e_X$ is the phase error rate and thus, $h(e_X)$ is the cost of performing privacy amplification in order to remove the correlations between the corrected sifted key and Eve. The term $e_Z$ is the bit error rate and $fh(e_Z)$ corresponds to the amount of syndrome information required to make Alice's and Bob's keys the same. The quantities $Y_Z$ and $e_Z$ in Eq. (\ref{eq:keyrate}) can be directly obtained from an implementation of the experiment. Therefore, we are left with the estimation of the phase error rate. We do this below. 

\subsection{Estimation of the phase error rate}
We assume that Alice prepares the states $\ket{\Phi_{j\beta}}_{BE}$ as defined in Eq. (\ref{eq:multi-mode}). These states take into consideration the non-qubit assumption, a possible THA by Eve, and SPFs. In the virtual protocol (see Appendix \ref{app:security} for further details), Alice prepares the following state in the $Z$ basis:
\begin{linenomath*}
\begin{equation} 
\ket{\Psi_{Z}}_{ABE} = \frac{1}{\sqrt{2}} \Big[\ket{0_Z}_A \otimes \big( a_{0Z} \ket{\phi_{0Z}}_{BE}+ b_{0Z} \ket{\phi^{\perp}_{0Z}}_{BE}\big) + \ket{1_Z}_A \otimes \big( a_{1Z} \ket{\phi_{1Z}}_{BE} +  b_{1Z} \ket{\phi^{\perp}_{1Z}}_{BE}\big) \Big].
\label{eq:stateZ}
\end{equation}\end{linenomath*}
We then define the bit error rate as 
\begin{linenomath*}
\begin{equation}
e_{Z} = \frac{Y_{0Z,1Z}^{(Z)} + Y_{1Z,0Z}^{(Z)}}{Y_{0Z,0Z}^{(Z)} + Y_{1Z,0Z}^{(Z)} + Y_{0Z,1Z}^{(Z)} + Y_{1Z,1Z}^{(Z)}},
\label{eq:biterror2}
\end{equation}\end{linenomath*}
where the yields $Y_{sZ,jZ}^{(Z)}$, with $s,j \in \{0,1\}$, are the joint probabilities that Alice prepares the state $\ket{\Psi_{Z}}_{ABE}$, Bob selects the $Z$ basis, and Alice (Bob) obtains the bit value $j$ ($s$) when she (he) measures the system $A$ $(B)$ in the $Z$ basis. Note that, the superscript $(Z)$ in the yields represents the basis used in the state preparation while the subscripts denote the bases employed in the measurements. These yields are directly observed in the experiment. Similarly, the phase error rate is defined as
\begin{linenomath*}
\begin{equation}
e_X = \frac{Y_{0X,1X}^{(Z) vir} + Y_{1X,0X}^{(Z) vir}}{Y_{0X,0X}^{(Z) vir} + Y_{1X,0X}^{(Z) vir} + Y_{0X,1X}^{(Z) vir} + Y_{1X,1X}^{(Z) vir}},
\label{eq:ex}
\end{equation}\end{linenomath*}
where $Y_{sX,jX}^{(Z) vir}$, with $s,j \in \{0,1\}$, is the joint probability that Alice prepares the state $\ket{\Psi_{Z}}_{ABE}$, she and Bob select the $Z$ basis but both use the $X$ basis for their measurements (rather than the selected $Z$ basis), and Alice (Bob) obtains the bit value $j$ ($s$). The phase error rate corresponds to the bit error in the virtual protocol. Also, we have that the denominator of $e_X$ in Eq.(\ref{eq:ex}) is equal to $Y_{0Z,0Z}^{(Z)} + Y_{1Z,0Z}^{(Z)} + Y_{0Z,1Z}^{(Z)} + Y_{1Z,1Z}^{(Z)}$, since by assumption the probability to obtain an inconclusive outcome associated to the operator $\hat{M}_f$ is the same for the both basis for any incoming state. This means that to estimate $e_X$ we only need to calculate the virtual yields $Y_{0X,1X}^{(Z)vir}$ and $Y_{1X,0X}^{(Z)vir}$. 

In the virtual protocol, after Alice measures the system $A$ in Eq. (\ref{eq:stateZ}) in the $X$ basis, she sends Bob the (unnormalised) states: 
\begin{linenomath*}\begin{equation}
\hat{\theta}_{BE, jX, vir} = \Tr_A \big[ \dyad{j_X}{j_X}_A \otimes \hat{\mathds{1}}_{BE} \dyad{\Psi_{Z}}{\Psi_{Z}}_{ABE} \big],
\label{eq:density_matrix}
\end{equation}\end{linenomath*}
where $\Tr_A$ is the partial trace over the virtual system $A$. Using Eqs. (\ref{eq:stateZ}) and (\ref{eq:density_matrix}) we can calculate the unnormalised states sent by Alice for $j \in \{0,1\}$ and obtain that $\hat{\theta}_{BE, jX, vir} =  \dyad{\psi}{\psi}_{BE, jX, vir}$ with
\begin{linenomath*}\begin{equation}
\ket{\psi}_{BE,jX,vir} = \frac{1}{2} \Big[ a_{0Z} \ket{\phi_{0Z}}_{BE} + b_{0Z} \ket{\phi^{\perp}_{0Z}}_{BE} + (-1)^j \big(a_{1Z} \ket{\phi_{1Z}}_{BE} +  b_{1Z} \ket{\phi^{\perp}_{1Z}}_{BE}\big) \Big].
\label{eq:unstate}
\end{equation}\end{linenomath*}
Writing Eq. (\ref{eq:unstate}) in terms of the states $\ket{\gamma_{jX}}_{BE}$ and $ \ket{\gamma_{jX}^\perp}_{BE}$, defined below, we have that:
\begin{linenomath*}\begin{equation}
\begin{split}
\ket{\psi}_{BE, jX, vir} = \frac{1}{2}  \bigg[&\sqrt{|a_{0Z}|^2 + (-1)^j \big( a_{0Z}^* a_{1Z}{\braket{\phi_{0Z}|\phi_{1Z}}}_{BE} + a_{0Z} a_{1Z}^* {\braket{\phi_{1Z}|\phi_{0Z}}}_{BE}  \big) + |a_{1Z}|^2} ~  \ket{\gamma_{jX}}_{BE} \\
&+ \sqrt{|b_{0Z}|^2 + (-1)^j \big( b_{0Z}^* b_{1Z}{\braket{\phi_{0Z}^\perp | \phi_{1Z}^\perp}}_{BE} + b_{0Z} b_{1Z}^* {\braket{\phi_{1Z}^\perp | \phi_{0Z}^\perp}}_{BE}  \big) + |b_{1Z}|^2}  ~ \ket{\gamma_{jX}^\perp}_{BE}\bigg],
\label{eq:statej}
\end{split}
\end{equation}\end{linenomath*}
where the normalised states $\ket{\gamma_{jX}}_{BE}$ have the form 
\begin{linenomath*}\begin{equation}
\ket{\gamma_{jX}}_{BE} = \frac{a_{0Z}\ket{\phi_{0Z}}_{BE}  + (-1)^j a_{1Z} \ket{\phi_{1Z}}_{BE}}{ \sqrt{|a_{0Z}|^2 + (-1)^j \big(a_{0Z}^* a_{1Z} {\braket{\phi_{0Z} | \phi_{1Z}}}_{BE} + a_{0Z} a_{1Z}^* {\braket{\phi_{1Z}|\phi_{0Z}}}_{BE}  \big) + |a_{1Z}|^2}}, 
\end{equation}\end{linenomath*}
and the normalised states $ \ket{\gamma_{jX}^\perp}_{BE}$, which are orthogonal to $\ket{\gamma_{jX}}_{BE} $, are given by
\begin{linenomath*}\begin{equation}
\ket{\gamma_{jX}^\perp}_{BE} = \frac{b_{0Z}\ket{\phi_{0Z}^\perp}_{BE}  + (-1)^j b_{1Z} \ket{\phi_{1Z}^\perp}_{BE}}{ \sqrt{|b_{0Z}|^2 + (-1)^j \big(b_{0Z}^* b_{1Z}  {\braket{\phi_{0Z}^\perp | \phi_{1Z}^\perp}}_{BE} + b_{0Z} b_{1Z}^*  {\braket{\phi_{1Z}^\perp | \phi_{0Z}^\perp}}_{BE}  \big) + |b_{1Z}|^2}}.
\end{equation}\end{linenomath*}
Note that, in Eq. (\ref{eq:statej}), we have decomposed $\ket{\psi}_{BE,jX,vir}$ into a single-mode qubit $\ket{\gamma_{jX}}_{BE}$ and a state in any mode orthogonal to it, $ \ket{\gamma_{jX}^\perp}_{BE}$. This decomposition follows the definition provided in Eq. (\ref{eq:multi-mode}), and it is an essential step for our estimation of the phase error rate.

To obtain the yields $Y_{sX,jX}^{(Z)vir}$ we need to calculate
\begin{linenomath*}\begin{equation}
Y_{sX,jX}^{(Z) vir} = P_{Z_A} P_{Z_B} \Tr[\hat{D}_{sX}\hat{\theta}_{BE, jX, vir} ],
\label{eq:yield}
\end{equation}\end{linenomath*}
where $\hat{D}_{sX} = \sum_k \hat{A}_k^{\dagger} \hat{M}_{sX} \hat{A}_k $ corresponds to Eve's action, represented by the Kraus operators $\hat{A}_k$, as well as Bob's measurement with $\hat{M}_{sX}$ being an element of Bob's POVM. Here, recall the definition of the phase error rate where the $Z$ basis is selected but both Alice and Bob use the $X$ basis for their measurements (rather than the selected $Z$ basis), which is why $P_{Z_A}$ and $P_{Z_B}$ appear in Eq. (\ref{eq:yield}). Moreover, here we assume, for simplicity, that Eve applies the same quantum operation to every signal, which corresponds to a collective attack, but our analysis can be generalised to coherent attacks by considering the Azuma's inequality \cite{azuma} (see Appendix \ref{app:security}), which deals with any correlations among the events, {\it {\it i.e.}}, the phase error rate pattern. Using Eqs. (\ref{eq:statej})-(\ref{eq:yield}) we obtain the following expression for the yields: 
\begin{linenomath*}\begin{equation}
\begin{split}
Y_{sX,jX}^{(Z)vir} =  &  ~P_{Z_A} P_{Z_B} \bigg( A_j \Tr\Big[\hat{D}_{sX}  \dyad{\gamma_{jX}}{\gamma_{jX}}_{BE}  \Big]  \\ 
& + \Tr\Big[\hat{D}_{sX}\Big(B_j \dyad{\gamma_{jX}}{\gamma_{jX}^{\bot}}_{BE} + B_j^* \dyad{\gamma_{jX}^{\bot}}{\gamma_{jX}}_{BE} + C_j \dyad{\gamma_{jX}^{\bot}}{\gamma_{jX}^{\bot}}_{BE}\Big)\Big]\bigg),
\label{eq:yield1}
\end{split}
\end{equation}\end{linenomath*}
where the coefficients $A_j$, $B_j$, and $C_j$ are defined in Appendix \ref{app:coefficients}, and we omit presenting their explicit expressions here for simplicity. Since the state $\ket{\gamma_{jX}}_{BE}$ in the first term of Eq. (\ref{eq:yield1}) is a single-mode qubit state, its density matrix can be expressed as 
\begin{linenomath*}\begin{equation}
\hat{\rho}_{jX} =   \dyad{\gamma_{jX}}{\gamma_{jX}}_{BE}  = \frac{1}{2} \sum_i P^{jX,vir}_i \hat{\sigma_i},
\end{equation}\end{linenomath*}
where $P^{jX,vir}_i$ are the coefficients of the Bloch vector and $\hat{\sigma_i}$, with $i \in \{Id,x,y,z\}$, represent the identity and the three Pauli operators, respectively. Therefore, we have that 
\begin{linenomath*}\begin{equation}
A_j  \Tr\Big[\hat{D}_{sX}  \dyad{\gamma_{jX}}{\gamma_{jX}}_{BE}  \Big] = A_j \big[P^{jX,vir}_{Id} q_{sX|Id} + P^{jX,vir}_x q_{sX|x} + P^{jX,vir}_y q_{sX|y} + P^{jX,vir}_z q_{sX|z}\big], 
\end{equation}\end{linenomath*}
where $P^{jX,vir}_i = \Tr \big[\hat{\sigma_i}  \dyad{\gamma_{jX}}{\gamma_{jX}}_{BE} \big]$ and $q_{sX|i} = \frac{1}{2} \Tr \big[\hat{D}_{sX} \hat{\sigma}_i \big]$ can be regarded as the transmission rates of the operator $\hat{\sigma}_i$. These can be calculated by solving a system of linear equations with the events from the actual protocol, which we will explain later. Moreover, by choosing the $y$ axis of the Bloch sphere appropriately we can always set $P^{jX,vir}_y =0$ for all the Bloch vectors, since the PM just creates rotations in the $X$-$Z$ plane of the Bloch sphere. Indeed, even if the PM introduces loss depending on Alice's state selection, as long as the three states form a triangle in the Bloch sphere, we can apply such simplification \cite{tamaki}. As already mentioned in Section \ref{sec:description}, we note that any implementation of the loss-tolerant protocol requires that the three states form a triangle in the Bloch sphere.

Furthermore, it is possible to find both lower and upper bounds on the second term of Eq. (\ref{eq:yield1}). In particular, this term can be written as $\Tr[\hat{D}_{sX} N_j]$ where $N_j$ is the matrix $\Big[\begin{smallmatrix} C_j&B_j^*\\ B_j&0 \end{smallmatrix}\Big]$ with eigenvalues
\begin{linenomath*}\begin{equation}
\lambda_{max_j} = \frac{C_j + \sqrt{{C_j}^2 + 4|B_j|^2}}{2} ~~~\text{and}~~~\lambda_{min_j} = \frac{C_j - \sqrt{{C_j}^2 + 4|B_j|^2}}{2}.
\end{equation}\end{linenomath*}
Using the properties of POVMs we have that the operators $\hat{D}_{sX}$ have eigenvalues between 0 and 1, therefore $\Tr[\hat{D}_{sX} N_j]$ is bounded by $\lambda_{min_{j}} \le \Tr[\hat{D}_{sX} N_j] \le \lambda_{max_{j}}$, since $\lambda_{min_{j}}$ is negative. 

This means that the virtual yields satisfy: 
\begin{linenomath*}\begin{equation}
\begin{split}
P_{Z_A} P_{Z_B} \Big(A_j \big[q_{sX|Id} + P_x^{jX,vir} ~q_{sX|x} + P_z^{jX,vir} ~ q_{sX|z}\big] + \lambda_{min_{j}}\Big) \le ~Y_{sX,jX}^{(Z)vir} \\
 \le P_{Z_A} P_{Z_B} \Big(A_j \big[q_{sX|Id} + P_x^{jX,vir} ~ q_{sX|x} + P_z^{jX,vir}~ q_{sX|z}\big] + \lambda_{max_{j}}\Big). \\
\end{split}
\label{eq:coef2}
\end{equation}\end{linenomath*}

To find the transmission rates $q_{sX|i}$, the actual events we need to consider are those associated with the yields $Y_{sX,0Z}^{(Z)}$, $Y_{sX,1Z}^{(Z)}$ and $Y_{sX,0X}^{(X)}$. These are defined as $Y_{sX,j\beta}^{(\beta)} = P_{j\beta} P_{X_B} \Tr \big[\hat{D}_{sX} \dyad{\Phi_{j\beta}}{\Phi_{j\beta}}_{BE}\big]$ for $j\beta \in \{ 0Z, 1Z, 0X\}$, where the normalised actual states $\ket{\Phi_{j\beta}}_{BE}$ are defined in Eq. (\ref{eq:multi-mode}).
That is, $\ket{\Phi_{j\beta}}_{BE}$ are the states emitted by Alice in the actual protocol when she chooses the bit value $j$ and the basis $\beta$, in the presence of multi-mode signals. Using exactly the same method explained above, we obtain
\begin{linenomath*}\begin{equation}
\begin{split}
Y_{sX,j\beta}^{(\beta)} = & ~ P_{j\beta} P_{X_B}\bigg( E_{j\beta} \Tr\Big[\hat{D}_{sX}  \dyad{\phi_{j\beta}}{\phi_{j\beta}}_{BE} \Big] \\ 
&+ \Tr\Big[\hat{D}_{sX}\Big(F_{j\beta} \dyad{\phi_{j\beta}}{\phi_{j\beta}^{\bot}}_{BE} +F^*_{j\beta} \dyad{\phi_{j\beta}^{\bot}}{\phi_{j\beta}}_{BE} + G_{j\beta} \dyad{\phi_{j\beta}^{\bot}}{\phi_{j\beta}^{\bot}}_{BE}\Big)\Big]\bigg),
\end{split}
\label{eq:actualY}
\end{equation}\end{linenomath*}
where $E_{j\beta} = |a_{j\beta}|^2$, $F_{j\beta} =  a_{j\beta}b_{j\beta}^*$, $F^*_{j\beta} =  a_{j\beta}^*b_{j\beta}$ and $G_{j\beta} = |b_{j\beta}|^2$. Therefore, we find that the actual yields satisfy 
\begin{linenomath*}\begin{equation}
\begin{split}
P_{j\beta} P_{X_B} \Big(E_{j\beta} ~\big[q_{sX|Id} + P^{j\beta}_x q_{sX|x} + P^{j\beta}_z q_{sX|z}\big] + \lambda_{min_{j\beta}}\Big) \le ~ Y_{sX,j\beta}^{(\beta)} \\
 \le P_{j\beta} P_{X_B} \Big(E_{j\beta}~ \big[q_{sX|Id} + P^{j\beta}_x q_{sX|x} + P^{j\beta}_z q_{sX|z}\big] + \lambda_{max_{j\beta}}\Big), \\
\end{split}
\label{eq:hello}
\end{equation}\end{linenomath*}
where 
\begin{linenomath*}\begin{equation}
\lambda_{max_{j\beta}} = \frac{G_{j\beta} + \sqrt{G_{j\beta}^2 + 4|F_{j\beta}|^2}}{2}  ~~~\text{and}~~~ \lambda_{min_{j\beta}} = \frac{G_{j\beta} - \sqrt{G_{j\beta}^2 + 4|F_{j\beta}|^2}}{2},
\label{eq:coef3}
\end{equation}\end{linenomath*}
are the eigenvalues for the non-qubit part of the actual states and, $P^{j\beta}_{x}$ and $P^{j\beta}_{z}$ are the coefficients of the Bloch vector for the actual states. By substituting $j\beta \in \{ 0Z, 1Z, 0X\}$ in Eqs. (\ref{eq:hello}) and (\ref{eq:coef3}), we obtain a system of three linear inequalities, which can be expressed as 
\begin{linenomath*}\begin{equation}
\begin{split}
\begin{bmatrix} q_{sX|Id},&q_{sX|x},&q_{sX|z}\end{bmatrix} \hat{A} + \begin{bmatrix}  \lambda_{min_{0Z}},& \lambda_{min_{1Z}},& \lambda_{min_{0X}} \end{bmatrix} \le \begin{bmatrix} \frac{Y_{sX,0Z}^{(Z)}}{P_{0Z} P_{X_B}}, & \frac{Y_{sX,1Z}^{(Z)}}{P_{1Z} P_{X_B}}, & \frac{Y_{sX,0X}^{(X)}}{P_{0X} P_{X_B}} \end{bmatrix} \\
 \le  \begin{bmatrix} q_{sX|Id},&q_{sX|x},&q_{sX|z}\end{bmatrix} \hat{A} + \begin{bmatrix} \lambda_{max_{0Z}},&\lambda_{max_{1Z}},&\lambda_{max_{0X}} \end{bmatrix},
\end{split}
\label{eq:bounds}
\end{equation}\end{linenomath*} 
where $\hat{A} := (V^T_{0Z}, V^T_{1Z}, V^T_{0X})$ in which $V_{j\beta} = E_{j\beta} (1, P^{j\beta}_x,P^{j\beta}_z)$ and where the superscript $T$ means transpose. 
By rearranging Eq. (\ref{eq:bounds}) we obtain the bounds on the transmission rates $q_{sX|Id}$, $q_{sX|x}$, and $q_{sX|z}$ to be 
\begin{linenomath*}\begin{equation}
\begin{split}
\bigg( \begin{bmatrix} \frac{Y_{sX,0Z}^{(Z)}}{P_{0Z} P_{X_B}}, & \frac{Y_{sX,1Z}^{(Z)}}{P_{1Z} P_{X_B}}, & \frac{Y_{sX,0X}^{(X)}}{P_{0X} P_{X_B}}\end{bmatrix} - \begin{bmatrix} \lambda_{max_{0Z}},&\lambda_{max_{1Z}},&\lambda_{max_{0X}} \end{bmatrix} \bigg) \hat{{A}^{-1}} \le \begin{bmatrix} q_{sX|Id},&q_{sX|x}, &q_{sX|z}\end{bmatrix} \\
\le \bigg(\begin{bmatrix} \frac{Y_{sX,0Z}^{(Z)}}{P_{0Z} P_{X_B}}, & \frac{Y_{sX,1Z}^{(Z)}}{P_{1Z} P_{X_B}}, & \frac{Y_{sX,0X}^{(X)}}{P_{0X} P_{X_B}} \end{bmatrix} - \begin{bmatrix}  \lambda_{min_{0Z}},& \lambda_{min_{1Z}},& \lambda_{min_{0X}} \end{bmatrix} \bigg) \hat{{A}^{-1}},
\end{split}
\label{eq:bounds2}
\end{equation}\end{linenomath*}
where $\hat{A}^{-1}$ is the inverse of the matrix $\hat{A}$. 

By solving Eq. (\ref{eq:bounds2}), we can calculate the transmission rates and then substitute them into Eq. (\ref{eq:coef2}) to find the upper bounds on the virtual yields $Y_{0X,1X}^{(Z)vir}$ and $Y_{1X,0X}^{(Z)vir}$. Finally, by using these upper bounds on the virtual yields and the yields from the actual events we can estimate the phase error rate $e_X$ in Eq. (\ref{eq:ex}). 

As already mentioned previously, this technique is quite general and could be applied to many other QKD protocols. As an example, in Appendix \ref{app:mdi} we outline how this analysis could be performed for MDI-QKD.

\section{Simulation of the key rate}
\label{sec:simulation}
\subsection{Particular device model}
Only for the purpose of the simulation, we now consider a particular device model and a particular THA. In general, to experimentally guarantee that the three states emitted by Alice remain in two dimensions, i.e., in a single-mode qubit, her PM needs to have the same temporal, spectral, spatial and polarisation mode independently of the bit and basis choices. However, due to imperfections in the devices this condition is hard to fulfil. Some counter-measures against these imperfections have been suggested \cite{xu,xu2,jiang,mynbaev}, but they cannot rigorously ensure a single-mode qubit. Therefore, it is crucial to consider how device's flaws can be taken into account in a security proof. This is the aim of our analysis. For simplicity, among many imperfections, we select the polarisation mode as an example of how to use our framework. 

A change in polarisation can arise from the imperfect alignment of the laser with the principal axis of the PM and/or when the PM is polarisation dependent, i.e., the state of polarisation of the signals prepared might be different for each encoding phase value. In principle, this could be avoided by using a polarisation beam splitter (PBS) that selects a single polarisation mode. In practice, however, because of the finite extinction ratio of the PBS this is usually not the case. Here, we relax the need for a perfect PBS by considering a polarisation multi-mode scenario. We remark, nonetheless, that our analysis can be applied to any multi-mode scenario. Using our formalism, we can express the states sent by Alice in the scenario considered in an analogous way to Eq. (\ref{eq:multi-mode}):
\begin{equation}
 \ket{\Omega_{j\beta}}_B = \cos \theta_{j\beta} \ket{\omega_{j\beta}}_{HB}+ \sin \theta_{j\beta} \ket{\omega_{j\beta}}_{VB},
\label{eq:polarisation}
\end{equation}
for $j\beta \in \{0Z,1Z,0X\}$, where the subscripts $H$ and $V$ refer to the horizontal and vertical polarisation modes, respectively. That is, now the polarisation state of $\ket{\omega_{j\beta}}_B$ depends on Alice's bit and basis choices instead of being the same independently of her encoding. Next, we add the SPF and the THA to this particular device model.

For the states $\ket{\omega_{j\beta}}_{HB}$ and $ \ket{\omega_{j\beta}}_{VB}$ we use the definitions in Eq. (\ref{eq:alice_states}), where they both live in a qubit space. Also, by using Eq. (\ref{eq:alice_states}) these states already include SPFs whenever the parameter $\delta > 0$. As stressed in Section \ref{sec:description}, since in this case we know the form of the states we do not need to consider the worst case scenario but only the inner product$\tensor[_{HB}]{\braket{\omega_{j\beta} | \omega_{j'\beta'}}}{_{VB}}= 0$ for all $j,j',\beta$ and $\beta'$.

Additionally, we consider an active information leakage in our device model. For this, we assume that Eve sends strong light into Alice's PM, which is then back-reflected and exits Alice's lab in the form 
\begin{equation}
\ket{\xi_{j\beta}}_E = C_I \ket{e}_E + C_D \ket{e_{j\beta}}_E.
\label{eq:eve_states}
\end{equation}
In this expression, $|C_I|^2 + |C_D|^2 = 1$, and $\ket{e}_E$ ($\ket{e_{j\beta}}_E$) represents (represent) the setting independent (dependent) state (states) on Alice's bit and basis choice, where we assume that ${\braket{e | e_{j\beta}}}{_E} = 0$. That is, the state $\ket{e}_E$ ($\ket{e_{j\beta}}_E$) provides Eve with no (some) information about Alice's bit and basis values each given time. Therefore, our model for the THA can be parameterised by only two parameters, $C_I$ and $C_D$, and no further detailed information is needed to apply our analysis. For instance, when we increase isolation on Alice's sending device, the independent component increases and Eve obtains less information about the states being sent. Moreover, in the absence of further information about the states $\ket{e_{j\beta}}_E$, we assume the worst case scenario where these states are orthogonal to each other, i.e., ${\braket{e_{j\beta} | e_{j'\beta'}}}{_E} = 0$ for any $(j,\beta) \neq (j',\beta')$. Clearly, if Alice and Bob know the states $\ket{e_{j\beta}}_E$, this information can be trivially included in the formalism below.

If $\ket{\xi_{j\beta}}_E$ is say, for instance, a coherent state, $\ket{e}_E$ is the vacuum state (i.e., $\ket{e}_E = \ket{v}_E$), $C_I = e^{-\mu/2}$ and $C_D = \sqrt{1 - e^{-\mu}}$, where $\mu$ is the intensity of Eve's back reflected light. In this case, note that the condition ${\braket{e_{j\beta} | e_{j'\beta'}}}{_E} = 0$ is not satisfied since Eve will never be able to perfectly distinguish the states dependent on Alice's encoding. The value of this overlap depends on the isolation of the devices. Below, however, we conservatively assume for simplicity the worst case scenario where this overlap is zero. 

Putting Eq. (\ref{eq:polarisation}) and Eq. (\ref{eq:eve_states}) together, Alice's emitted state for the single photon pulses is modelled as 
\begin{equation}
\ket{\Phi_{j\beta}}_{BE} = \ket{\Omega_{j\beta}}_B \otimes \ket{\xi_{j\beta}}_E. 
\label{eq:psi}
\end{equation}
By using Eqs. (\ref{eq:alice_states})-(\ref{eq:polarisation})-(\ref{eq:eve_states})-(\ref{eq:psi}) and by assuming that $\ket{\xi_{j\beta}}_E$ are coherent states, we obtain
\begin{equation}
\begin{split}
\ket{\Phi_{j\beta}}_{BE} &= \big( \cos \theta_{j\beta} \ket{\omega_{j\beta}}_{HB} + \sin \theta_{j\beta} \ket{\omega_{j\beta}}_{VB} \big) \otimes \big(C_I \ket{v}_E + C_D \ket{e_{j\beta}}_E \big) \\
& = \cos \theta_{j\beta} C_I  \ket{\omega_{j\beta}}_{HB} \ket{v}_E + \cos \theta_{j\beta} C_D \ket{\omega_{j\beta}}_{HB} \ket{e_{j\beta}}_E  + \sin \theta_{j\beta} \ket{\omega_{j\beta}}_{VB} \otimes  \big(C_I \ket{v}_E + C_D \ket{e_{j\beta}}_E \big). \\
\end{split}
\label{eq:fullstate}
\end{equation}
The first term of Eq. (\ref{eq:fullstate}) has polarisation $H$ and is insensitive to the THA, it corresponds to $a_{j\beta} \ket{\phi_{j\beta}}_{BE}$ in Eq. (\ref{eq:multi-mode}). Similarly, the other terms have either polarisation $V$ and/or are affected by the THA, and together they correspond to $b_{j\beta} \ket{\phi_{j\beta}^\perp}_{BE}$ in Eq. (\ref{eq:multi-mode}). In this case, the unnormalised virtual states given by Eq. (\ref{eq:statej}) have now the form
%
%
\begin{equation}
\begin{split}
\ket{\psi}_{BE, jX, vir} = \frac{1}{2}  \Bigg[& C_I \sqrt{\cos \theta_{0Z}^2 - (-1)^j ~2\cos \theta_{0Z} \cos \theta_{1Z} \sin \frac{\delta}{2} + \cos \theta_{1Z}^2} ~  \ket{\gamma_{jX}}_{BE} \\
&+ \sqrt{C_I^2 \bigg(\sin \theta_{0Z}^2 - (-1)^j ~2 \sin \theta_{0Z} \sin \theta_{1Z} \sin \frac{\delta}{2} + \sin \theta_{1Z}^2 \bigg) + 2C_D^2 }  ~ \ket{\gamma_{jX}^\perp}_{BE}\Bigg],
\label{eq:statejj}
\end{split}
\end{equation}
where we have used the relationship ${\braket{\omega_{0Z} | \omega_{1Z}}}{_B}={\braket{\omega_{0Z} | \omega_{1Z}}}{_B}= - \sin (\frac{\delta}{2})$. In order to estimate the phase error rate, we need to calculate the transmission rates $q_{sX|Id}$, $q_{sX|x}$ and $q_{sX|z}$ using the actual yields. For this, we use Eq. (\ref{eq:bounds2}) where in this particular example, the matrix $\hat{A}$ is 
\begin{equation}
\hat{A} = \begin{bmatrix}
E_{0Z}&E_{1Z}&E_{0X}\\ 0 & -E_{1Z}\sin(\delta) & E_{0X}\sin(\pi/2 + \delta/2) \\ E_{0Z} & -E_{1Z}\cos(\delta) & E_{0X}\cos(\pi/2 + \delta/2) 
\end{bmatrix},
\end{equation}
where $E_{j\beta} = C_I^2 \cos^2 \theta_{j\beta}$. Then, we can find the virtual yields by using Eq. (\ref{eq:coef2}) where, in this example, the coefficients of the Bloch vectors are 
\begin{equation}
\begin{split}
& P_x^{jX,vir} = \frac{(-1)^j ~2 \cos \theta_{0Z} \cos \theta_{1Z} \cos \frac{\delta}{2} - 2 \cos \theta_{1Z}^2  \cos \frac{\delta}{2}  \sin \frac{\delta}{2}}{\cos \theta_{0Z}^2 - (-1)^j ~2 \cos \theta_{0Z} \cos \theta_{1Z} \sin \frac{\delta}{2} + \cos \theta_{1Z}^2}, \\
& P_z^{jX,vir} = \frac{\cos \theta_{0Z}^2 - (-1)^j ~2 \cos \theta_{0Z} \cos \theta_{1Z} \sin \frac{\delta}{2} + \cos \theta_{1Z}^2 \big(1 - 2 \cos^2 \frac{\delta}{2}\big)}{\cos \theta_{0Z}^2 - (-1)^j ~2 \cos \theta_{0Z} \cos \theta_{1Z} \sin \frac{\delta}{2} + \cos \theta_{1Z}^2}. \\
\end{split}
\end{equation}
Finally, one can directly use Eq. (\ref{eq:ex}) to estimate the phase error rate $e_X$.  

\subsection{Lo-Preskill's analysis with imperfectly characterised states}
With the method described above, it is possible to employ the leaky source without compromising the security of the QKD system. Nonetheless, depending on the situation and the particular experimental parameters it might be beneficial to consider another method that in some cases might provide a higher key generation rate. Therefore, it is important to compare our generalised loss-tolerant protocol with an alternative method, say the Lo-Preskill's analysis introduced in \cite{lo4}. In Appendix \ref{app:lopreskill}, we provide a detailed description of this analysis.

To ensure a fair comparison between both protocols we consider the efficient four-state loss-tolerant protocol, where the four states of the BB84 protocol \cite{bennett} are used to run two loss-tolerant protocols simultaneously. That is, when Alice emits the state $\ket{\omega_{0X}}_B$ ($\ket{\omega_{1X}}_B$) she considers that it belongs to the first (second) loss-tolerant protocol, while each of the two protocols is randomly chosen by Alice before sending the pulse. See Eq. (\ref{eq:extrastate}) for the definition of the state $\ket{\omega_{1X}}_B$ in the presence of SPFs. This means that no modifications to the hardware of the standard BB84 protocol are required and therefore, the four-state loss-tolerant protocol is equivalent to the BB84 protocol from an experimental point of view. Furthermore, we consider the same assumptions about the states for both of them. Namely, we employ the decomposition of the state of the single-photon pulses emitted by Alice into a single-mode qubit and any other modes which are orthogonal to the former, i.e., Eq. (\ref{eq:multi-mode}). For the simulations, we apply Lo-Preskill's analysis to the same particular device model described in Section \ref{sec:simulation}.A.

In order to quantify the phase error rate $e_X$ in the Lo-Preskill's analysis we use the following expression \cite{lo4}
\begin{equation}
e_X  \le {e_Z} + 4 \Delta' (1-\Delta') (1 - 2 {e_Z}) + 4 (1 - 2\Delta' ) \sqrt{\Delta' (1 - \Delta') {e_Z} (1 - {e_Z})},
\label{eq:eph_delta}
\end{equation}
which depends on the bit error rate $e_Z$ and on the imbalance $\Delta'$ of a quantum coin. To find this imbalance, we need to calculate the inner product of Eqs. (\ref{eq:ent1}) and (\ref{eq:ent2}). In turn, these equations depend on the states $\ket{\Phi_{j\beta}}_{BE}$, defined in Eq. (\ref{eq:fullstate}), therefore, their overlaps allows us to find $\Delta'$. Using Eq. (\ref{eq:fullstate}) we can calculate the inner product of these states to be
\begin{equation}
{\braket{\Phi_{j\beta} | \Phi_{j'\beta'}}}{_{BE}} = \cos \theta_{j\beta} \cos \theta_{j'\beta'} C_I^2 {\braket{\omega_{j\beta} | \omega_{j'\beta'}}}{_B},
\label{eq:overlap2}
 \end{equation} 
where $j,j' \in \{0,1\}$ and $\beta,\beta' \in \{X,Y\}$. 

Using the same definition for the secret key rate given by Eq. (\ref{eq:keyrate}), below we compare both security approaches for the same device model as a function of the device parameters.

\subsection{Results and discussion}
We show the results obtained for $R$ as a function of the overall system loss (which includes both the channel attenuation and the loss at Bob's receiver), for different values of $\delta$, $\theta_{j\beta}$ and $\mu$ which correspond to the SPFs, non-qubit assumption and THA respectively. The angles $\theta_{j\beta}$ are chosen such that they are associated with Alice's encoding of the states $\ket{\omega_{j\beta}}_{B}$. That is, $\theta_{0Z}=0$, $\theta_{1Z}=\pi \hat\theta$ and $\theta_{0X}=\frac{\pi}{2}\hat\theta$ for a certain angle $\hat\theta$. In our simulations, we consider the experimental parameters to be: the dark count rate $p_d = 10^{-7}$, $f = 1.16$ and the fiber loss coefficient $\alpha = 0.2$ dB/Km. Moreover, we assume for simplicity that in the loss-tolerant protocol and in the Lo-Preskill's analysis $P_{Z_A} = P_{Z_B} = \frac{1}{2}$. This selection of probabilities might not be ideal but it is sufficient for the purpose of the simulation. 
By using the channel model described in Appendix \ref{app:channel}, we find that $Y_Z = Y_{0Z,0Z}^{(Z)} + Y_{1Z,0Z}^{(Z)} + Y_{0Z,1Z}^{(Z)} + Y_{1Z,1Z}^{(Z)} = P_{Z_A} P_{Z_B} \big[ 4 (1 - \frac{\eta}{2}) p_d + \eta\big]$ where $\eta$ is the overall transmission efficiency of the system. The bit error rate is then given by
\begin{linenomath*}\begin{equation}
e_Z = \frac{2\big(1-\frac{\eta}{2}\big) p_d + \frac{\eta}{2} + \frac{\eta}{4}(\cos 2 \delta + \cos \delta)(p_d-1)}{4\big(1-\frac{\eta}{2}\big) p_d + \eta}.
\label{eq:biterror}
\end{equation}\end{linenomath*}

\subsubsection{Generalised loss-tolerant protocol}
In order to evaluate how the different imperfections of the source affect the key generation rate we analyse each of them separately. For the simulations, we select the SPFs according to the experimental results reported in \cite{xu,honjo,li}. There are some works related with the mode dependency \cite{xu,tang2}, but unfortunately they do not directly provide the value of $\hat\theta$. Therefore, we evaluate $\hat\theta$ over a big range and choose the angles $\theta_{j\beta}$ to be: $\theta_{0Z}=0$, $\theta_{1Z}=\pi \hat\theta$ and $\theta_{0X}=\frac{\pi}{2}\hat\theta$ for a certain angle $\hat\theta$. This choice comes from Alice's encoding of the different states, which means that $\theta_{j\beta}$ is associated with the prepared state $\ket{\omega_{j\beta}}_{B}$. Obviously, a better experimental characterisation of the source would be essential to improve the accuracy of the current parameters. Finally, for the intensity of Eve's back reflected light during the THA, we select a range based on the work presented in \cite{lucamarini}. 

Furthermore, we consider both the cases when the mode dependency parameter $\theta_{j\beta}$, which is associated with the non-qubit assumption, is independent and dependent on Alice's bit and basis choice. When it is independent (i.e., when $\theta_{j\beta} = \theta$), Eqs. (\ref{eq:fullstate}) and (\ref{eq:statejj}) are simplified and become, respectively 
\begin{equation}
\begin{split}
\ket{\Phi_{j\beta}}_{BE} &= \big( \cos \theta \ket{\omega_{j\beta}}_{HB} + \sin \theta \ket{\omega_{j\beta}}_{VB} \big) \otimes \big(C_I \ket{v}_E + C_D \ket{e_{j\beta}}_E \big) \\
& = \cos \theta~ C_I  \ket{\omega_{j\beta}}_{HB} \ket{v}_E + \big[ \cos \theta~ C_D \ket{\omega_{j\beta}}_{HB} \ket{e_{j\beta}}_E  + \sin \theta \ket{\omega_{j\beta}}_{VB} \otimes  \big(C_I \ket{v}_E + C_D \ket{e_{j\beta}}_E \big) \big], \\
\end{split}
\label{eq:fullstate2}
\end{equation}
and
\begin{equation}
\ket{\psi}_{BE,jX,vir} = C_I \cos \theta \sqrt{2 - (-1)^j ~2\sin \frac{\delta}{2}} ~\ket{\gamma_{jX}}_{BE} +  \sqrt{C_I^2 \sin^2 \theta \bigg(2 - (-1)^j~ 2\sin \frac{\delta}{2}\bigg) + 2 C_D^2} ~\ket{\gamma_{jX}^\perp}_{BE}.
\end{equation} 
Using these equations, and following the method described in Section \ref{sec:estimation}, we estimate the phase error rate. When the mode dependency parameter is setting dependent, we use Eqs.  (\ref{eq:fullstate}) and (\ref{eq:statejj}) with $\theta_{j\beta}$, for $j\beta \in \{0Z,1Z,0X\}$.

Fig. \ref{fig:lossTolerant}(a) demonstrates that even if $\delta$ increases the key rate stays approximately the same, which means that Eve cannot enhance the flaws of the signals by exploiting channel loss. This is one main advantage of the loss-tolerant protocol \cite{tamaki}.

We consider the case when $\theta$ is independent of Alice's encoding in Fig. \ref{fig:lossTolerant}(b). The perfect scenario, i.e, when $\theta = 0$, implies that the signals prepared by Alice are in the single horizontal mode $H$, as seen in Eq. (\ref{eq:fullstate2}). If $\theta$ increases, the fraction of vertical polarisation also increases and the states sent becomes progressively more imperfect and vulnerable to a possible attack. For instance, these states form a three dimensional Hilbert space hence, Eve can perform an unambiguous state discrimination (USD) attack \cite{chefles, dusek}, in which she sends the identified state to Bob only when the USD measurement succeeds. This results in a decrease of the secret key rate as shown in Fig. \ref{fig:lossTolerant}(b). Additionally, our results show that as long as $\theta$ is sufficiently small (for the experimental parameters considered this means $\theta \lessapprox 10^{-7}$) the effect on the secret key rate is small, since it is approximately the same as when $\theta=0$. It is important to note, however, that because we are assuming a setting independent $\theta$ we have the freedom to choose a good polarisation mode. For instance, instead of the definition used above we could have called $\cos \theta \ket{\omega_{jZ}}_{HB}+\sin \theta \ket{\omega_{jZ}}_{VB}$ our mode if Alice had identified $\theta$ exactly, and regard it as the single-mode qubit, in turn affecting the results shown in Fig. \ref{fig:lossTolerant}(b). This means that Eve would be unable to exploit this flaw and obtain information about the key, and hence the secret key rate would be unaffected. This scenario would be equivalent to $\theta = 0$ and it is not the one considered in Fig. \ref{fig:lossTolerant}(b). For the setting dependent parameter $\theta_{j\beta}$, we obtain a slightly lower key rate as expected, since this scenario provides more information about Alice's encoding to a possible eavesdropper. The reason for this difference lies in the orthogonality of the states, which increases when we have a setting dependent $\theta_{j\beta}$. 

Finally, Fig. \ref{fig:lossTolerant}(d) shows how the THA affects the generalised loss-tolerant protocol, where $\mu$ quantifies the intensity of Eve's back reflected light. An increase in $\mu$ results in a lower secret key rate since Eve acquires more information about Alice's encoding thus compromising the security of the system. Furthermore, no key can be obtained around $\mu \gtrapprox 10^{-3}$. By comparing Figs. \ref{fig:lossTolerant}(b) and \ref{fig:lossTolerant}(d), it can be seen that there is a relationship between $\mu$ and $\theta$. In particular, the resulting secret key rate coincides when $\mu \sim \theta^2$. For example, the key rate is the same when $\mu = 10^{-6}$ and when $\theta= 10^{-3}$. This relationship can also be obtained analytically from the coefficient of the first term in Eq. (\ref{eq:fullstate2}), by using approximations to the series expansion of $\cos \theta$ and $e^{-\mu}$.

\subsubsection{Comparison between the generalised loss-tolerant protocol and the Lo-Preskill's analysis}
In a similar manner, we can evaluate how the secret key generation rate $R$ depends on the device parameters for the Lo-Preskill's analysis. The results and discussion are in Appendix \ref{app:lopreskill}. Below we compare both security proofs and identify which one provides a better $R$ depending on the experimental set-up. This way an experimentalist can choose which method to use for known device parameters, and ensure the security of the generated key between Alice and Bob. Here, we select the SPFs to be either $\delta = 0.063$ or $\delta = 0.126$, and for the mode dependency we choose $\hat\theta = 10^{-3}$ and $\hat\theta = 10^{-5}$. In this comparison we use the setting dependent mode dependency parameter since it corresponds to a more realistic scenario. Finally, for the intensity of Eve's back reflected light during the THA we use $\mu = 10^{-10}$, $\mu = 10^{-7}$ and $\mu = 10^{-4}$. The results are shown in Fig. \ref{fig:comparison1}. Note that, the blue and red dashed lines coincide (for the resolution presented) in all graphs. The reason lies in the value of the variable $\mu$. That is, for $\mu = 10^{-10}$ and $\mu = 10^{-7}$, the Lo-Preskill's analysis results in approximately the same secret key rate (see Appendix \ref{app:lopreskill} for more details).

\begin{figure}[H]
	\centering
	\begin{subfigure}[H]{0.45\textwidth}
	\includegraphics[width=7.85cm]{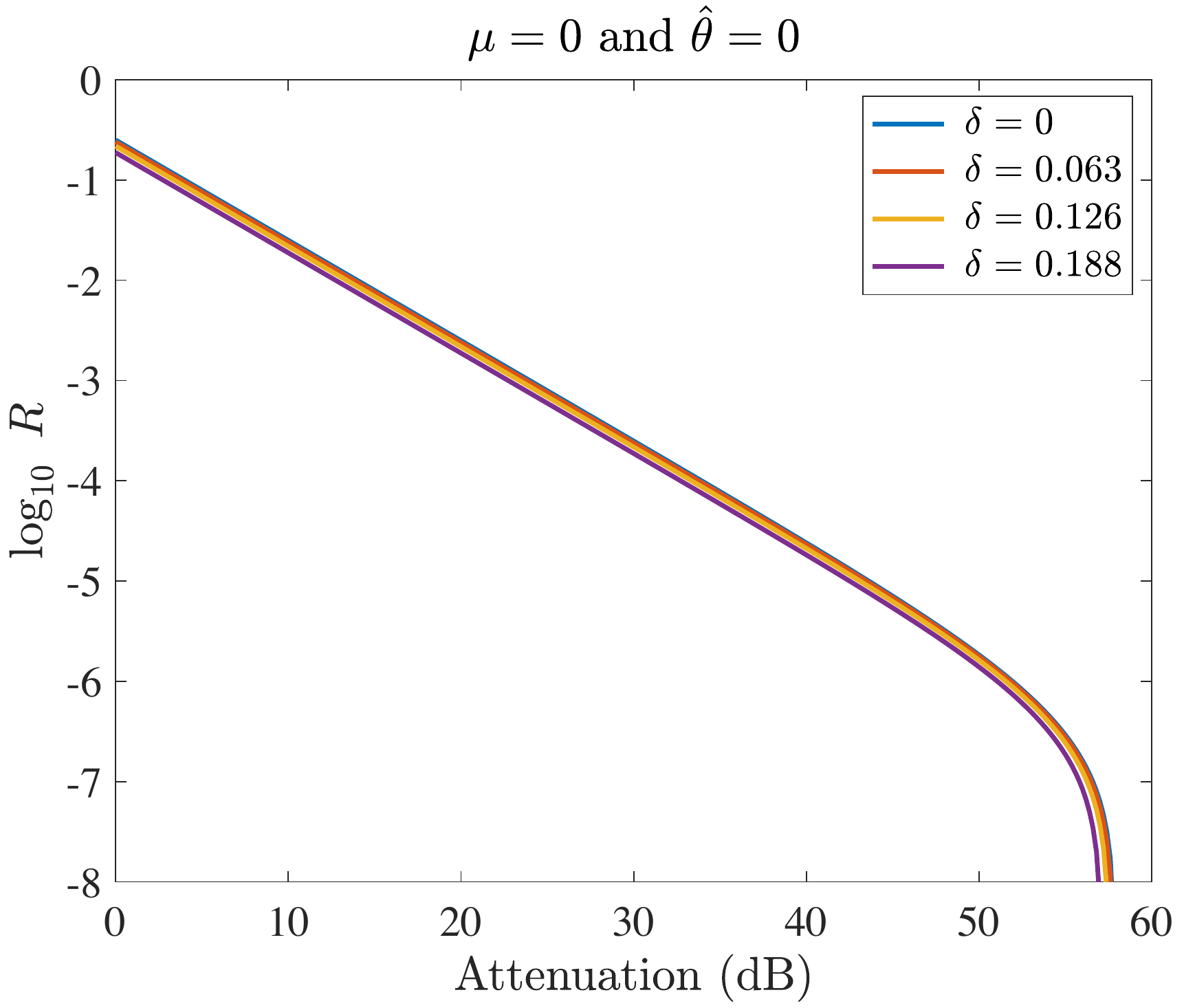} 
	\caption{}
	\end{subfigure}
	\begin{subfigure}[H]{0.45\textwidth}
	\includegraphics[width=7.85cm]{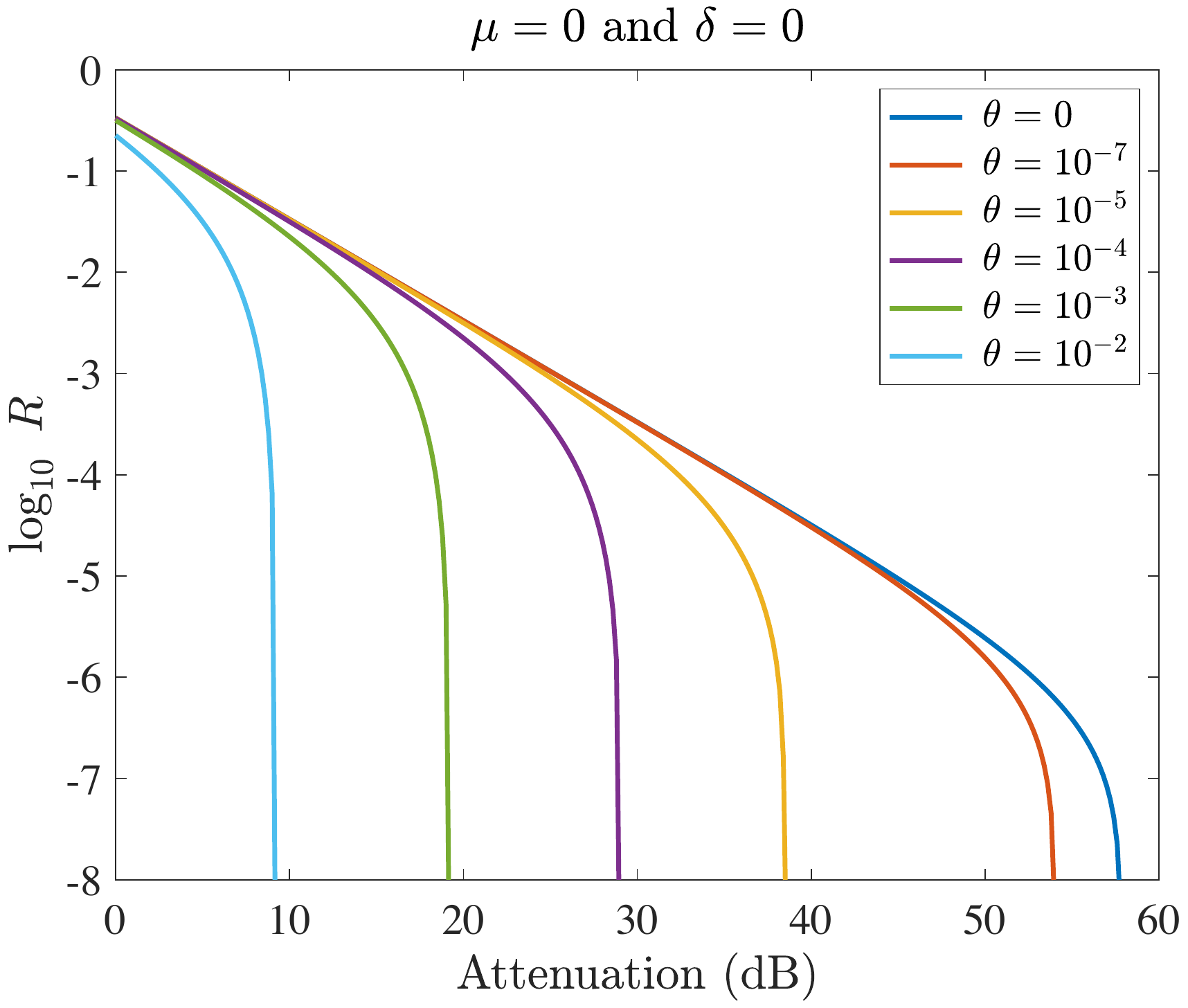}
	\caption{}
	\end{subfigure}
	\begin{subfigure}[H]{0.45\textwidth}
	\includegraphics[width=7.8cm]{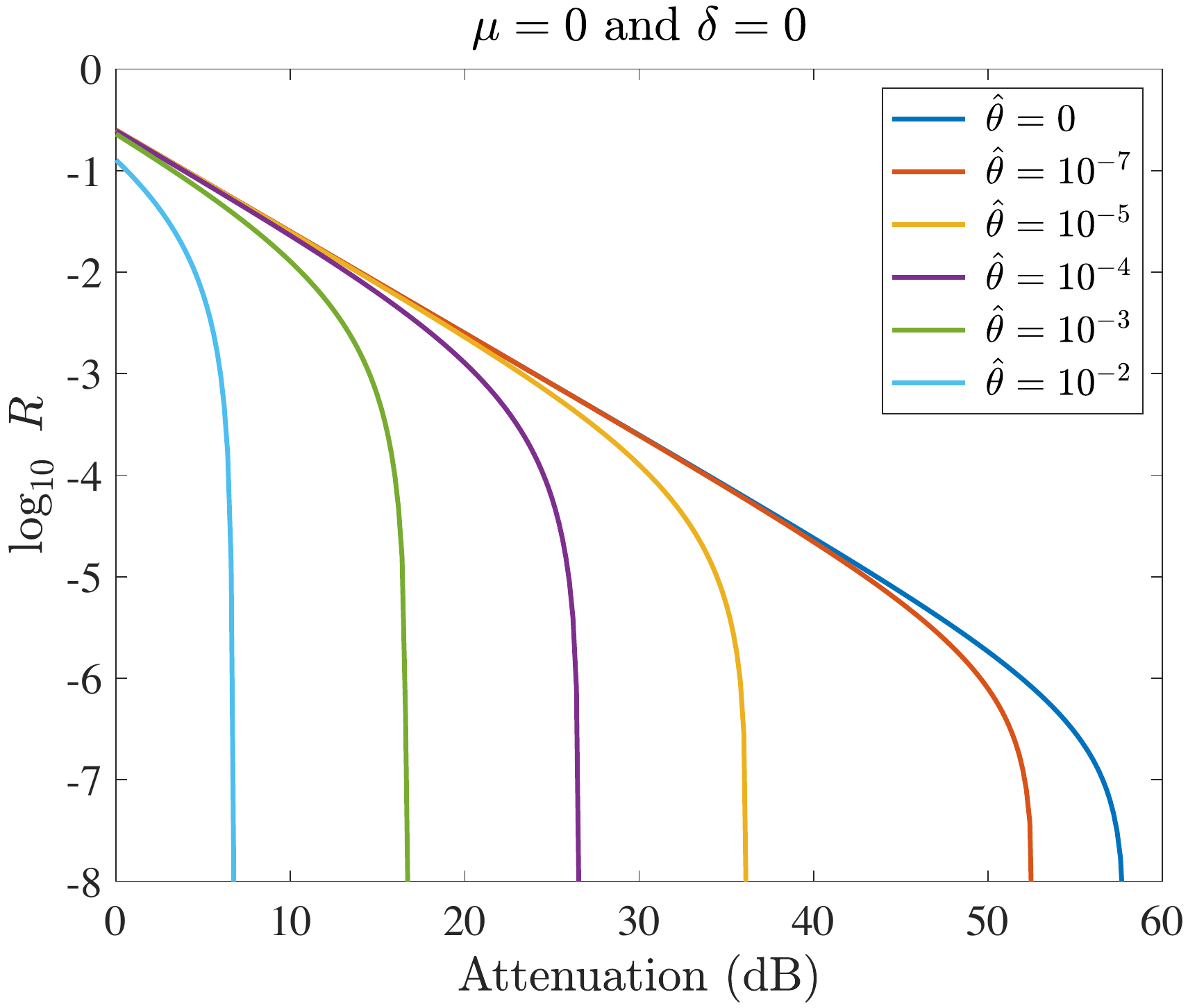}
	\caption{}
	\end{subfigure}
	\begin{subfigure}[H]{0.45\textwidth}
	\includegraphics[width=7.85cm]{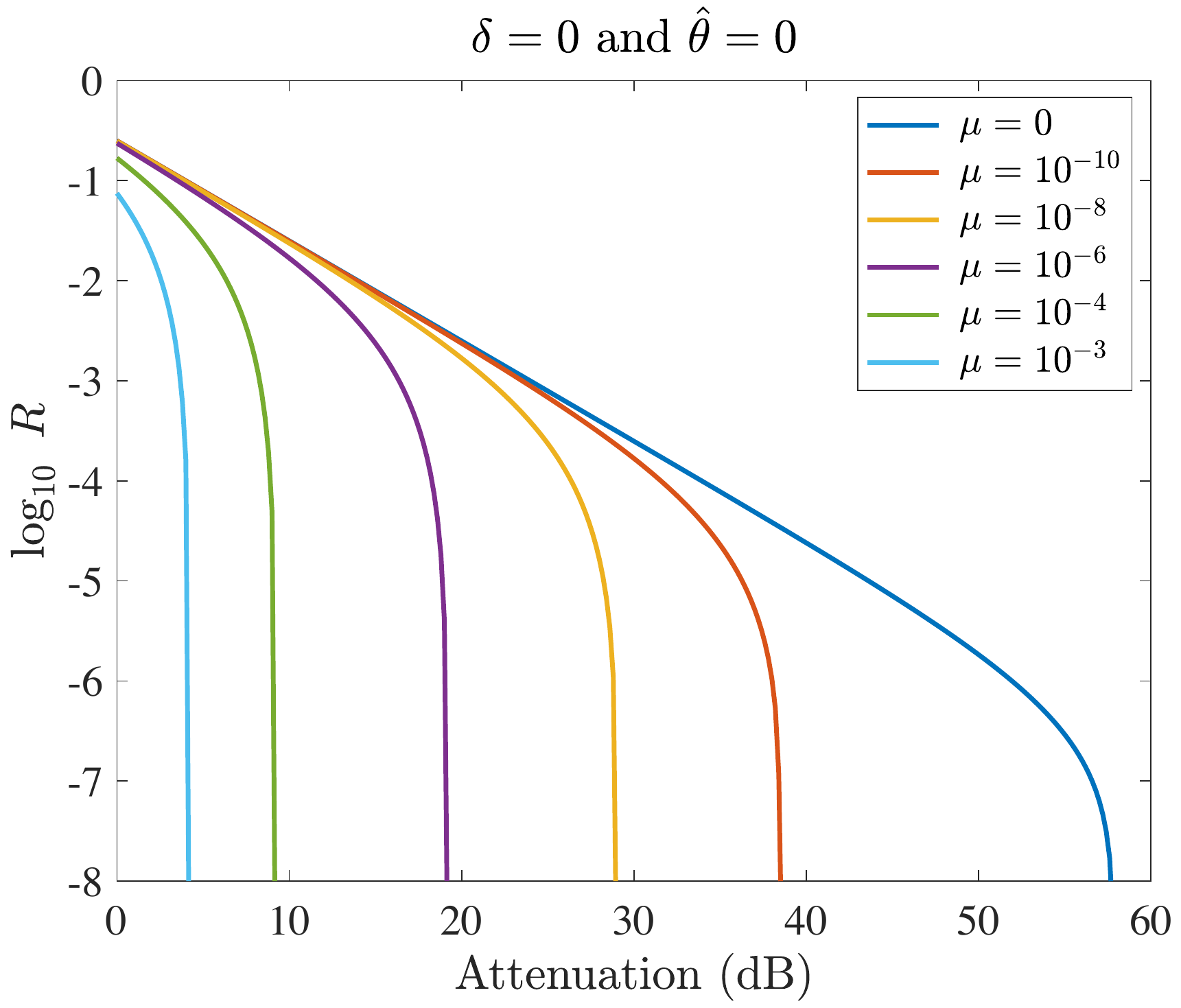}
	\caption{}
	\end{subfigure}
	\caption{Asymptotic secret key rate $R$ versus the overall system loss measured in dB for the generalised loss-tolerant protocol for various values of $\delta$, $\theta_{j\beta}$ and $\mu$. When we change the value of one of these parameters the others are kept constant and set to zero, which allows us to observe how each parameter affects the secret key rate. (a) Even if the parameter $\delta$ that characterises the SPFs increases, $R$ stays almost the same. (b) Setting independent $\theta$: As $\theta$ gets larger, the further away we are from the qubit scenario, hence $R$ decreases. (c) Setting dependent $\theta_{j\beta}$: The key generation rate decreases even further due to passive information leakage, in particular, $\hat\theta \gtrapprox 10^{-2}$ no longer provides a positive key generation rate. (d) As the intensity $\mu$ increases Eve obtains more information about the key causing the key rate to decrease.} 
	\label{fig:lossTolerant}
\end{figure}

By comparing Figs. \ref{fig:comparison1}(a) and \ref{fig:comparison1}(c), or Figs. \ref{fig:comparison1}(b) and \ref{fig:comparison1}(d), we can see how an increase in the parameter $\delta$, which is associated with SPFs, affects both protocols. For the generalised loss-tolerant protocol (LT) the key rate stays approximately the same as expected, since this method is loss tolerant to SPFs. On the other hand, the Lo-Preskill's analysis (LP) is more influenced by SPFs (see also Fig. \ref{fig:extLP} in Appendix \ref{app:lopreskill}). The reason for this difference is that in LP it is assumed the worst case scenario, in which Eve can enhance the basis dependence of the signals by exploiting the channel loss. However, in LT no such assumption is required hence the performance is maintained. This means that LT will typically outperform LP in the presence of high SPFs.

To compare LT and LP as a function of the setting dependent $\theta_{j\beta}$ we can contrast Figs. \ref{fig:comparison1}(a) and \ref{fig:comparison1}(b), or Figs. \ref{fig:comparison1}(c) and \ref{fig:comparison1}(d). The graphs show clear differences due to decreasing the value of $\hat\theta$, especially for the LT case. In Fig. \ref{fig:comparison1}(a) LP reaches a longer distance for any value of $\mu$, but when $\hat\theta = 10^{-5}$ LT gets better, particularly for $\mu = 10^{-10}$ as seen in Fig. \ref{fig:comparison1}(b). Furthermore, Fig. \ref{fig:comparison1}(b) shows that even when there are SPFs, LP can still do better \\
%
%
\begin{figure}[H]
	\centering
	\begin{subfigure}[H]{0.45\textwidth}
	\includegraphics[width=7.85cm]{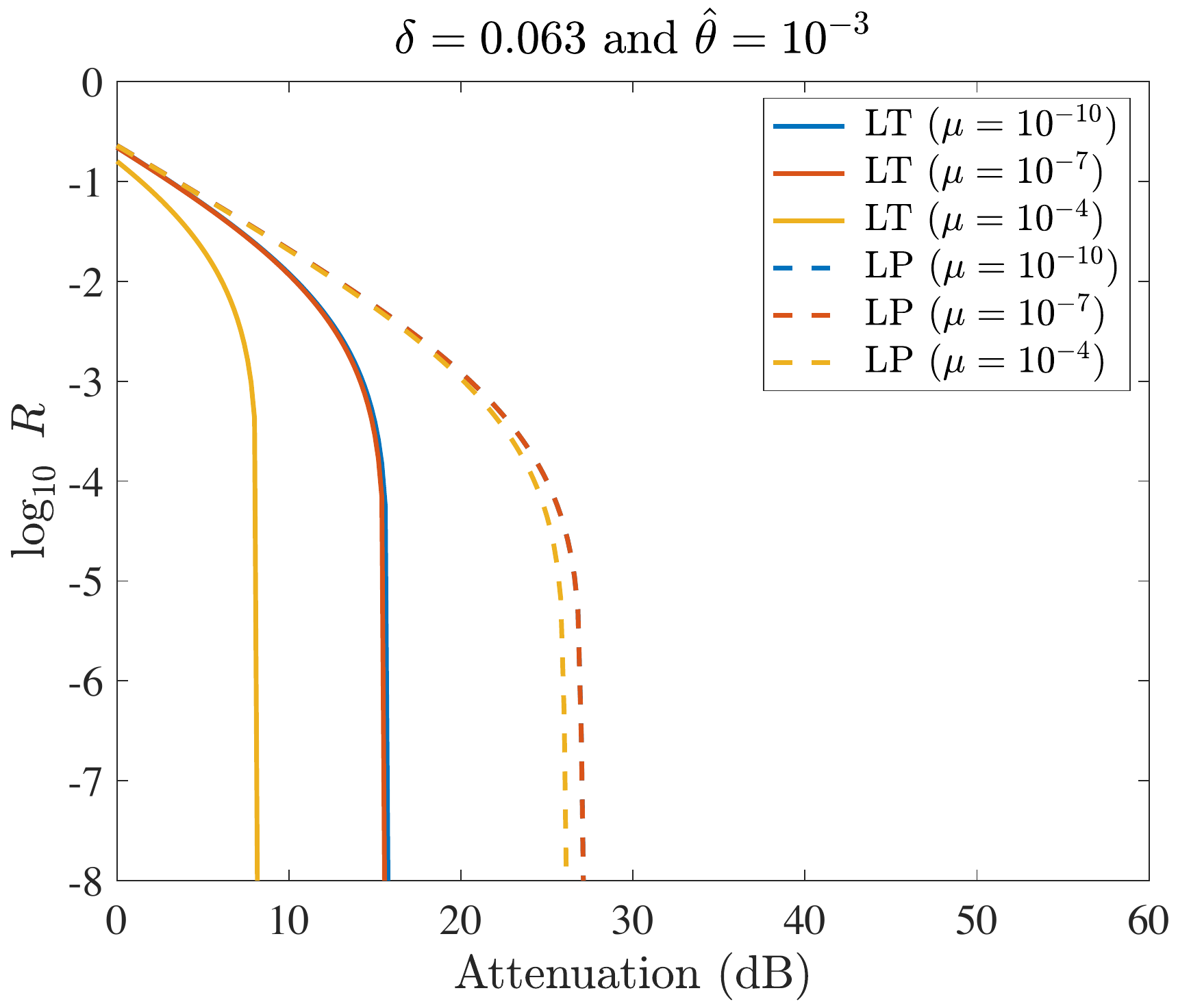} 
	\caption{}
	\end{subfigure}
	\begin{subfigure}[H]{0.45\textwidth}
	\includegraphics[width=7.85cm]{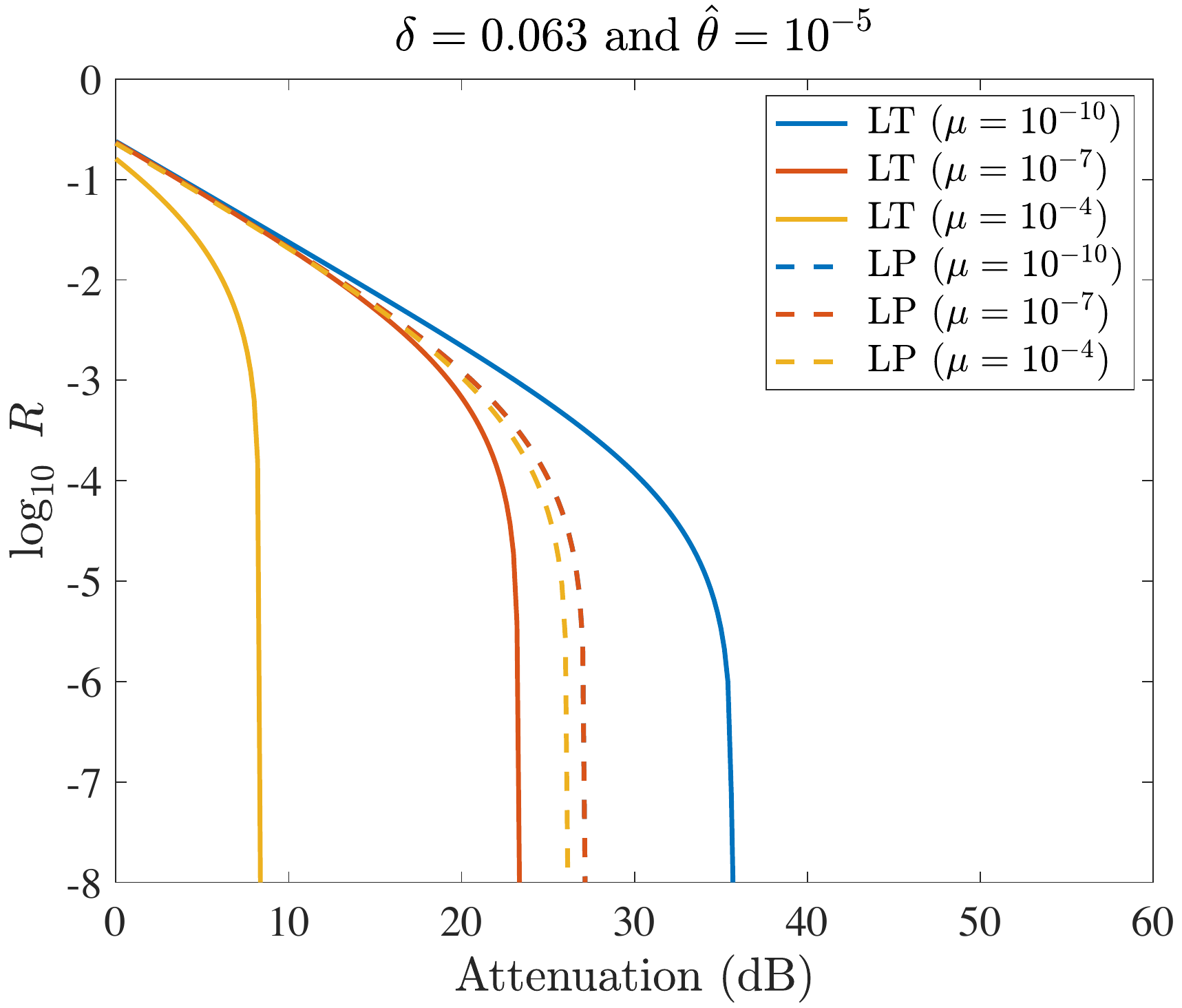}
	\caption{}
	\end{subfigure}
	\begin{subfigure}[H]{0.45\textwidth}
	\includegraphics[width=7.85cm]{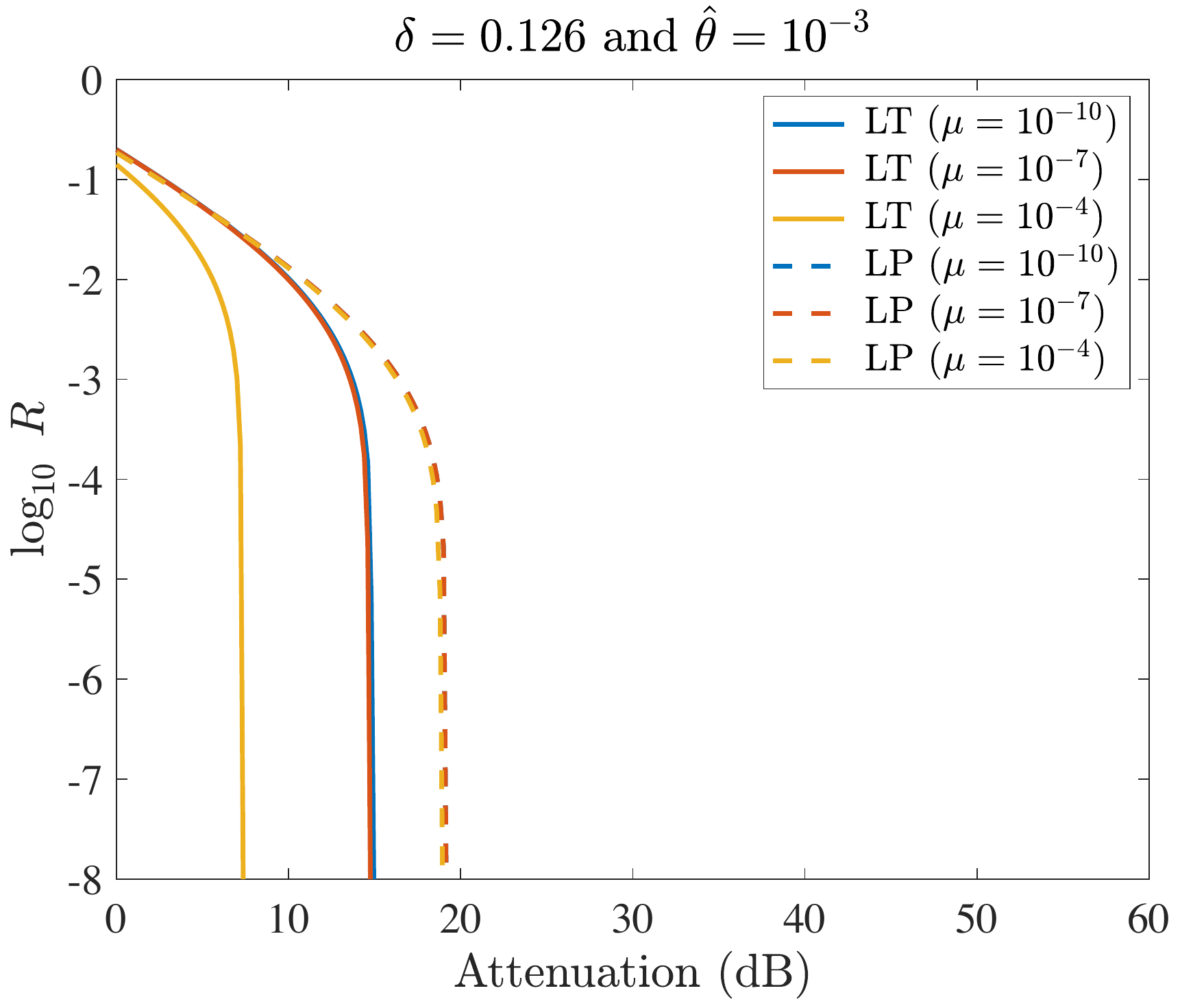}
	\caption{}
	\end{subfigure}
	\begin{subfigure}[H]{0.45\textwidth}
	\includegraphics[width=7.85cm]{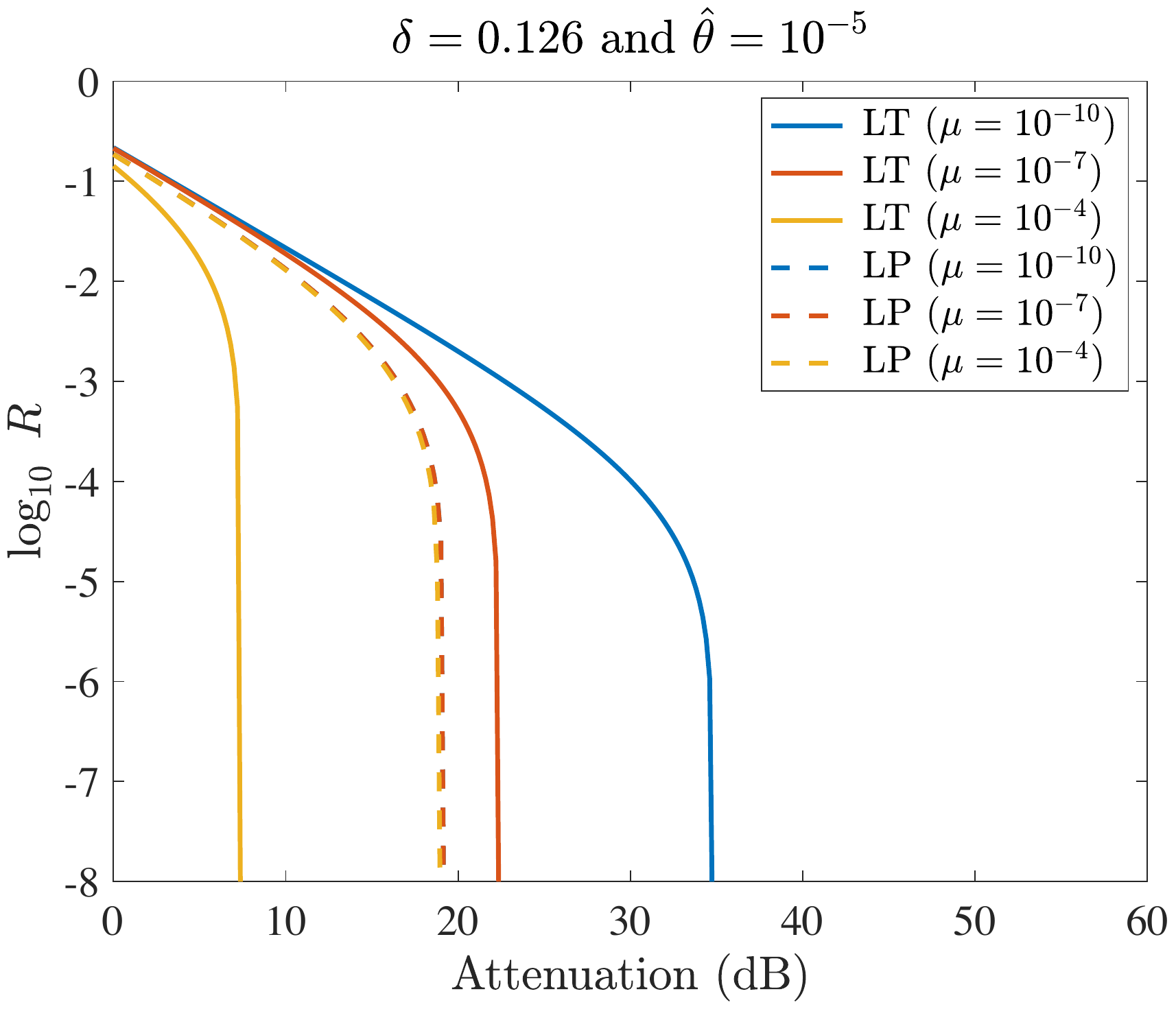} 
	\caption{}
	\end{subfigure}
	\caption{Secret key rate $R$ versus the overall system loss measured in dB for the generalised loss-tolerant protocol (LT) and Lo-Preskill (LP) analysis. The blue and red lines are superimposed in all graphs. (a) The Lo-Preskill's protocol performs better in this scenario because the SPFs are small but $\hat\theta$ is high. (b) For a smaller $\hat\theta$, the loss-tolerant analysis is better when $\mu= 10^{-10}$. (c) The generalised loss-tolerant protocol performs better when $\hat\theta$ is larger even if $\delta$ is high. (d) For large $\delta$ and small $\hat\theta$, the generalised loss-tolerant clearly surpasses the Lo-Preskill's analysis when $\mu= 10^{-10}$ or $\mu= 10^{-7}$.}
	\label{fig:comparison1}
\end{figure}
\noindent than LT if the states sent are far from the idealised qubit. This is because the non-qubit assumption negatively affects more LT than LP (see Figs. \ref{fig:lossTolerant}(c) and \ref{fig:extLP}(c)).

When we compare the values of $\mu$ for the LT and LP we can see a similar trend in the secret key rate for all graphs in Fig. \ref{fig:comparison1}. Namely, the difference between the curves when $\mu = 10^{-10}$ (blue) and $\mu = 10^{-4}$ (yellow) is much larger for LT than for LP, which means that the THA is worse for the LT. However, $\mu$ is a parameter that might be easily controlled experimentally by introducing passive countermeasures, such as optical isolators \cite{lucamarini}. Indeed, in \cite{lucamarini} it has been shown, for instance, that a value of $\mu = 10^{-6}$ could be easily achieved in practice. For example, even if Eve sends Alice optical pulses with $10^{20}$ photons, practical combinations of the components of Alice's transmitter could guarantee a total optical isolation of -170 dB, which would be enough to achieve $\mu = 10^{-6}$ \cite{lucamarini}. This means that the LT method may be a better alternative when the SPFs are more dominant and mode dependency is small, since it outperforms the LP analysis in Figs. \ref{fig:comparison1}(b) and \ref{fig:comparison1}(d). 

As explained above, the non-qubit assumption and the THA affect more the LT than the LP analysis. This might be because our generalisation of the loss-tolerant protocol is overestimating Eve. When we calculate the bounds for the yields we obtain that the eigenvalues $\lambda_{max}$ and $\lambda_{min}$ depend on the state preparation. However, this is probably too pessimistic because there might be some additional constraints among them, since the space spanned by the states associated to $0Z$ and $1Z$, respectively, is not orthogonal to the one spanned by the virtual states associated to $0X^{vir}$ and $1X^{vir}$. This means that these separate optimisations should not be possible in practise because Eve cannot achieve optimal values for all $\lambda$s. In other words, by improving our characterisation of the states we can improve the performance of the generalised loss-tolerant protocol. This is however beyond the scope of this paper and we leave it for future work. \\

\begin{figure}[H]
	\centering
	\begin{subfigure}[H]{0.45\textwidth}
	\includegraphics[width=7.8cm]{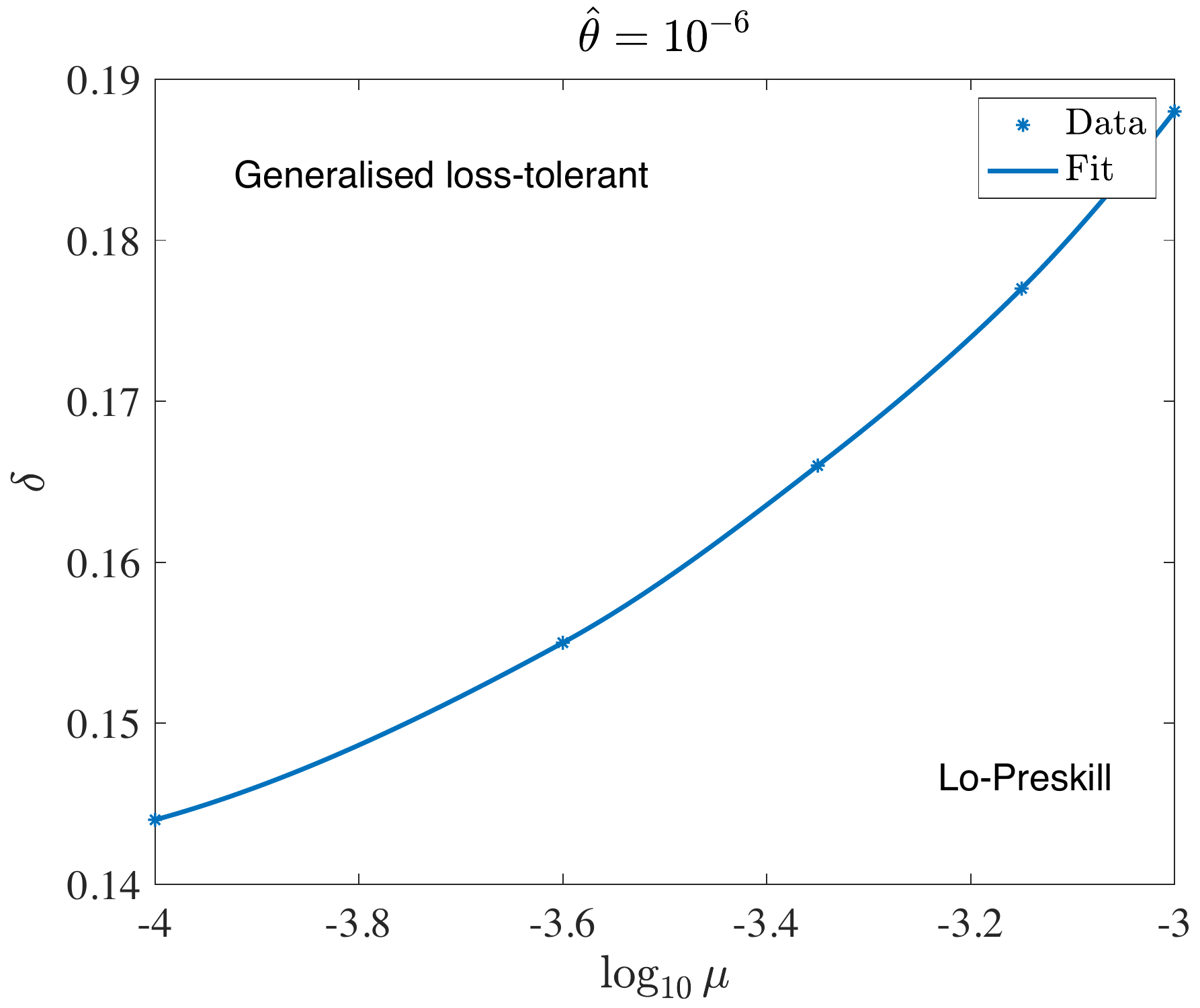}
	\caption{}
	\end{subfigure}
	\begin{subfigure}[H]{0.45\textwidth}
	\includegraphics[width=7.8cm]{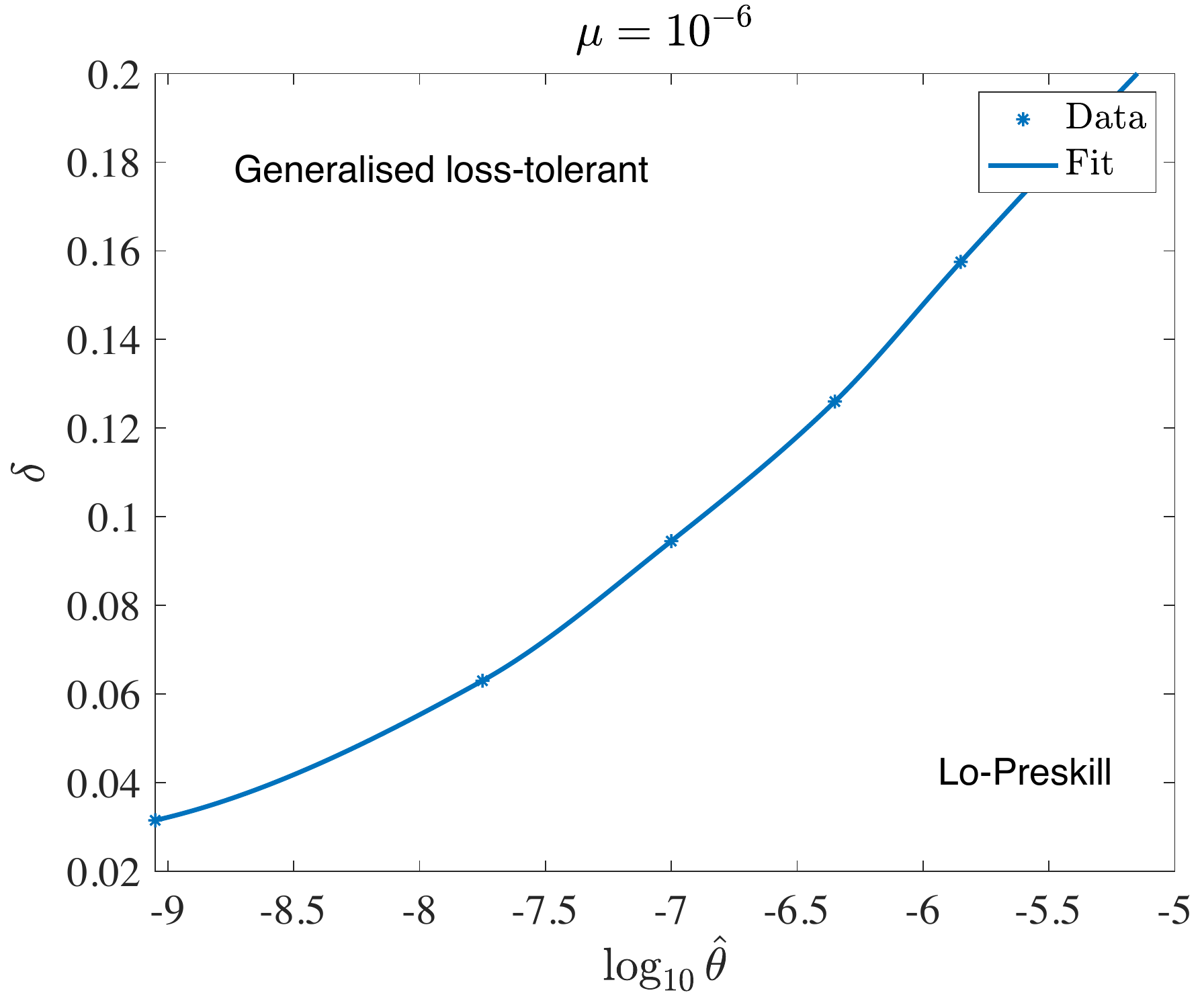}
	\caption{}
	\end{subfigure}
	\caption{The fitted line corresponds to those experimental parameters that result in the same key generation rate $R$ for both methods, the generalised loss-tolerant protocol and the Lo-Preskill's analysis. Above the line, the generalised loss-tolerant protocol performs better and below the line, the Lo-Preskill's analysis is the preferred method. The data points were fitted using a shape-preserving interpolant in Matlab. (a) Plot of $\delta$ against $\mu$ for $\hat\theta = 10^{-6}$. (b) Plot of $\delta$ against $\hat\theta$ for $\mu = 10^{-6}$.}
	\label{fig:comparison2}
\end{figure}

In order to further investigate the differences between the two methods, we determine the parameter regimes where their performance is identical. First, by setting $\hat\theta = 10^{-6}$, we can identify which values of $\delta$ and $\mu$ provide the same key generation rate $R$ for LT and LP. The results are presented in Fig. \ref{fig:comparison2}(a), where the diagram clearly shows which protocol performs better given a certain $\delta$ and $\mu$: above the fitted curve, the LT provides a better performance but below the curve LP is the preferable method. In other words, as the SPFs increase the LT is superior but, as $\mu$ increases LP becomes more suitable.

Similar results are obtained when $\mu = 10^{-6}$. This case is particularly useful since in principle we can control the value of $\mu$ experimentally by the amount of isolation we use in our devices. Again, as SPFs increase the LT becomes better, giving a better estimation of the phase error rate and a better secret key generation rate. 

\section{Conclusion}
\label{sec:conclusion}
Typical security proofs ignore many imperfections of experimental devices thus hindering the security claim of quantum key distribution (QKD). In this work, we have generalised the loss-tolerant QKD protocol to accommodate general imperfections. In particular, our formalism is valid for a general device model with, for instance, state preparation flaws (SPFs), mode dependency and Trojan horse attacks (THAs), which result in passive and active information leakage to an eavesdropper. Using this multi-mode scenario, we have shown that the qubit assumption can be removed from the loss-tolerant protocol without compromising the security of the QKD scheme. We present a formalism that can be used to estimate the phase error rate by finding the transmission rates of some virtual states and assuming the general state structure defined in Eq. (\ref{eq:multi-mode}). Therefore, in principle it can be applied to most QKD protocols. \\

In order to compare our generalised loss-tolerant protocol with other security proofs we have applied the Lo-Preskill's analysis \cite{lo4} to the same device model. In so doing, we have identified which approach delivers a higher secret key rate as a function of the experimental parameters. For example, the results obtained show that Lo-Preskill's method performs better under the non-qubit assumption and the THA but the generalised loss-tolerant protocol is better when there are SPFs. Since the THAs can be controlled using passive countermeasures, such as optical isolators, we have shown that in some cases the generalised loss-tolerant protocol might be the preferable method when the SPFs are more dominant. This way, our work can be used as a guideline to improve current experimental implementations in which multi-mode QKD is unavoidable. Moreover, it highlights the importance of source characterisation for more realistic security proofs. 

For completeness, we also note that Ref. \cite{wang3} has recently proposed a computational toolbox that can be used to numerically estimate the phase error rate of a QKD protocol, and such technique could be applied to the scenario considered in this paper. Essentially like the Lo-Preskill's analysis, their technique only requires the knowledge of the inner products between the states emitted by Alice and is mathematically simple, which is a striking difference to previous numerical analyses \cite{coles,winick}. That is, the approach in~\cite{wang3} can also remove the qubit assumption and include side-channels when estimating the phase error rate. There are, however, some relevant differences between that method and our formalism, besides the obvious one, i.e., that our work is an analytical technique. The approach in~\cite{wang3} requires a full characterisation of the side-channels in order to obtain the inner product of the states, while ours does not, resulting in a simpler characterisation of the source. Moreover, in the absence of side-channels, their method is not loss-tolerant in some parameter regimes, while ours is always loss-tolerant, which is essential to guarantee a good performance over long distances. Furthermore, their analysis considers pure states while our method also applies to the mixed-state scenario. Despite these differences, it would be interesting to combine the advantages of both methods to achieve a better implementation security, but we leave this for future works.

\section{Acknowledgements}
We thank Koji Azuma, Hoi-Kwong Lo, Go Kato, Norbert L{\"u}tkenhaus, Masato Koashi, Toshihiko Sasaki and Akihiro Mizutani, Guillermo Currás Lorenzo and Weilong Wang for very valuable discussions. This work was supported by the Spanish Ministry of Economy and Competitiveness (MINECO), the Fondo Europeo de Desarrollo Regional (FEDER) through grants TEC2014-54898-R and TEC2017-88243-R, and the European Union's Horizon 2020 research and innovation programme under the Marie Sk\l{}odowska-Curie grant agreement No 675662 (project QCALL). K.T. acknowledges support from JSPS KAKENHI Grant Numbers JP18H05237 18H05237, ImPACT Program of Council for Science, Technology and Innovation (Cabinet Office, Government of Japan), and JST-CREST JPMJCR 1671.


\appendix

\section{Three-state MDI-QKD}
\label{app:mdi}

Here, we describe how our generalised loss-tolerant protocol can be applied to MDI-QKD \cite{lo2}. We assume that Alice and Bob prepare three-states in the form of Eq. (\ref{eq:multi-mode}), thus including active and passive information leakage. These states are then sent to an untrusted relay Eve which is located between Alice and Bob. Note that, we name the relay Eve (instead of using the typical name of Charles) to simplify the notation, since in this way we can use the subscript E to denote all systems possessed by the eavesdropper. Eve is supposed to perform a Bell state measurement (BSM) and announce the results over a public channel. Alice and Bob keep the data associated with the successful events and discard the rest. Finally, to guarantee correctly correlated bit strings, either Alice or Bob apply a bit flip to part of their data. In this protocol, the $Z$ basis is used to generate the secret key and the $X$ basis for parameter estimation. 

As before, we can consider a virtual protocol in which Alice (Bob) prepares the state $\ket{\Psi_Z}_{ACE}$ ($\ket{\Psi_Z}_{BC'E}$) in the $Z$ basis, analogously to Eq. (\ref{eq:stateZ}), where $C$ ($C'$) is the system sent by Alice (Bob) to Eve. For simplicity, the discussion below considers the case where Eve's BSM only projects the incoming pulses into one Bell state: $\ket{\phi^+}$. In this case, the phase error rate is expressed as
\begin{equation}
e_X = \frac{Y_{0X,1X,\phi^+}^{(ZZ) vir} + Y_{1X,0X,\phi^+}^{(ZZ) vir}}{Y_{0X,0X,\phi^+}^{(ZZ) vir} + Y_{1X,0X,\phi^+}^{(ZZ) vir} + Y_{0X,1X,\phi^+}^{(ZZ) vir} + Y_{1X,1X,\phi^+}^{(ZZ) vir}},
\label{eq:ex_mdi}
\end{equation}
where $Y_{sX,jX,\phi^+}^{(ZZ) vir}$ with $s,j \in \{0,1\}$, is the joint probability that Alice (Bob) prepares the state $\ket{\Psi_Z}_{ACE}$ ($\ket{\Psi_Z}_{BC'E}$) and Eve declares the outcome $\ket{\phi^+}$, both Alice and Bob select the $Z$ basis but use the $X$ basis for their measurements of systems $A$ and $B$, and Alice (Bob) obtains the bit value $j$ ($s$). In order to find these virtual yields we need to calculate
\begin{equation}
Y_{sX,jX,\phi^+}^{(ZZ) vir} = P_{Z_A} P_{Z_B} \Tr[\hat{D}_{\phi^+} \hat{\theta}_{CE, jX, vir} \otimes \hat{\theta}_{C'E, sX, vir} ],
\label{eq:yield_mdi}
\end{equation}
where $\hat{D}_{\phi^+}$ corresponds to the announcement of Eve's outcome $\ket{\phi^+}$, and $\hat{\theta}_{CE, jX, vir}$ and $\hat{\theta}_{C'E, sX, vir}$ are defined similarly to those given by Eq. (\ref{eq:density_matrix}). Following a similar procedure to the one presented Section \ref{sec:estimation}, we can find the transmission rates to be $q_{\phi^+|ii'} = \frac{1}{4} \Tr [\hat{D}_{\phi^+} \hat{\sigma_i} \otimes \hat{\sigma_{i'}}]$ with $i$, $i' \in \{Id,x,y,z\}$. As in Eq. (\ref{eq:coef2}), we can calculate bounds of the virtual yields in terms of these quantities and their respective Bloch vector coefficients $P_i^{jX,vir}$.

Using the actual events $Y_{sZ,jZ,\phi^+}^{(ZZ)}$, $Y_{0X,jZ,\phi^+}^{(XZ)}$, $Y_{sZ,0X,\phi^+}^{(ZX)}$ and $Y_{0X,0X,\phi^+}^{(XX)}$ we can construct a system of nine linear inequalities. Finally, by employing the same assumption as before, {\it i.e.}, the three-states form a triangle in the Bloch sphere \cite{tamaki}, we can guarantee that these equations are linearly independent. Therefore, one can find bounds on the transmission rates, and consequently estimate the phase error rate.  

\section{Security proof against coherent attacks}
\label{app:security}

In this section, we present the security proof of our formalism against coherent attacks. For simplicity of the discussion, the main text deals with the case of pure states in a single-mode qubit space, however in this Appendix we consider the general scenario where the states could be mixed states in a single-mode qubit space. For this, we consider a virtual protocol \cite{shor, mayers}. This protocol is equivalent to the actual protocol in the sense that the resulting statistics of the measurements and the secret key rate generated between Alice and Bob are the same. Furthermore, the classical and quantum information available to Eve is equal in both protocols. The security claim follows from the fact that Alice and Bob can choose which protocol to execute and Eve is unable to distinguish between them. Hence, by proving the security of the virtual protocol we prove the security of the actual protocol. 

In this work we employ the complementary scenario \cite{koashi, koashi2}, which considers a virtual protocol that uses the complementary observable of the key generation basis. For instance, in the actual protocol Alice and Bob agree on the bit values in the $Z$ basis, while in the virtual protocol, they collaborate to prepare a qubit in an eigenstate of the $X$ basis. In doing so, the security proof basically reduces to the estimation of the phase error rate, which corresponds to the bit error rate that Alice and Bob would have observed if they would have measured the $Z$ basis state in the $X$ basis. Therefore, the aim of the virtual protocol is to estimate the phase error rate. In Section \ref{sec:estimation}, we showed how this can be done by using our formalism and how we can calculate the secret key rate $R$ against collective attacks. Here, we describe in detail the virtual protocol used for the security proof and explain how to accommodate coherent attacks by Eve through the use of Azuma's inequality \cite{azuma}. 
 
\subsection{Virtual protocol}

Here, we consider a more general case than that studied in the main text in which Alice generates a single-mode qubit system $B$, whose states are mixed states, and we show how to define the pure states needed for our security proof. We denote the mixed states by the density matrices $\hat{\rho}_{0Z_B}$, $\hat{\rho}_{1Z_B}$ and $\hat{\rho}_{0X_B}$. These states are diagonalised as
\begin{equation}
\begin{split}
&\hat{\rho}_{jZ_B} = P_{jZ}^{0} \dyad{\phi_{jZ}^{0}}_{B} + P_{jZ}^{1} \dyad{\phi_{jZ}^{1}}_{B}, \\
&\hat{\rho}_{0X_B} = P_{0X}^{0} \dyad{\phi_{0X}^{0}}_{B} + P_{jZ}^{1} \dyad{\phi_{0X}^{1}}_{B},
\end{split}
\end{equation}
where $j \in \{0,1\}$, and $P_{jZ}^0$, $P_{jZ}^1$, $P_{0X}^0$ and $P_{0X}^1$ are probabilities satisfying $P_{jZ}^0 + P_{jZ}^1 = 1$ and $P_{0X}^0 + P_{0X}^1 = 1$. Moreover, $\{\ket{\phi_{jZ}^0}_B, \ket{\phi_{jZ}^1}_B\}$ and $\{\ket{\phi_{0X}^0}_B, \ket{\phi_{0X}^1}_B\}$ are orthonormal bases in the single-mode qubit. The states sent might be mixed due to imperfections in Alice's devices, including a potential entanglement between her devices and Eve's ancilla. This means that in general these mixed states can be purified by introducing Alice's ancilla system $A_1$ and Eve's system $E$, and therefore we have the purifications of $\hat{\rho}_{0Z_B}$, $\hat{\rho}_{1Z_B}$ and $\hat{\rho}_{0X_B}$ as $\ket{\tilde{\psi}_{0Z}}_{A_1BE}$, $\ket{\tilde{\psi}_{1Z}}_{A_1BE}$, and $\ket{\tilde{\psi}_{0X}}_{A_1BE}$, each of which expressed by
\begin{equation}
\begin{split}
&\ket{\tilde{\psi}_{jZ}}_{A_1BE} = \sqrt{P_{jZ}^{0}} \ket{0_{jZ}}_{A_1E} \ket{\phi_{jZ}^{0}}_{B} + \sqrt{P_{jZ}^{1}} \ket{1_{jZ}}_{A_1E} \ket{\phi_{jZ}^{1}}_{B}, \\
&\ket{\tilde{\psi}_{0X}}_{A_1BE} = \sqrt{P_{0X}^{0}} \ket{0_{0X}}_{A_1E} \ket{\phi_{0X}^{0}}_{B} + \sqrt{P_{0X}^{1}} \ket{1_{0X}}_{A_1E} \ket{\phi_{0X}^{1}}_{B}.\\
\end{split}
\end{equation}
Here, $\{\ket{0_{jZ}}_{A_1E}, \ket{1_{jZ}}_{A_1E}\}$ and $\{\ket{0_{0X}}_{A_1E}, \ket{1_{0X}}_{A_1E}\}$ are orthonormal bases. Now, we define states similar to Eq. (\ref{eq:stateZ}) that include the purification of Alice's state: 
\begin{equation}
\begin{split}
&\ket{\tilde{\Psi}_{Z}}_{A_1A_2BE} = \frac{1}{\sqrt{2}} \Big[\ket{0_Z}_{A_2} \ket{\tilde{\psi}_{0Z}}_{A_1BE} + \ket{1_Z}_{A_2} \ket{\tilde{\psi}_{1Z}}_{A_1BE} \Big], \\
&\ket{\tilde{\Psi}_{X}}_{A_1A_2BE} = \ket{0_X}_{A_2} \ket{\tilde{\psi}_{0X}}_{A_1BE}, \\
\end{split}
\end{equation}
where $A_2$ is Alice's ancilla system used to generate a bit value in the protocol, i.e., it possesses information about Alice's encoding. As explained above, in the security analysis Alice measures $A_2$ in the $X$ basis instead of the $Z$ basis when $\ket{\tilde{\Psi}_{Z}}_{A_1A_2BE}$ is prepared, therefore, it is useful to write this state in the $X$ basis of system $A_2$. By substituting $\ket{0_Z}_{A_2} = \frac{1}{\sqrt{2}} (\ket{0_X}_{A_2} + \ket{1_X}_{A_2})$ and $\ket{1_Z}_{A_2} = \frac{1}{\sqrt{2}} (\ket{0_X}_{A_2} - \ket{1_X}_{A_2})$ we can express $\ket{\tilde{\Psi}_{Z}}_{A_1A_2BE}$ as 
\begin{equation}
\ket{\tilde{\Psi}_{Z}}_{A_1A_2BE} = \sqrt{\frac{1 + \braket{ \tilde{\psi}_{0Z} | \tilde{\psi}_{1Z}}_{A_1BE}}{2}} \ket{0_X}_{A_2} \ket{\tilde{\psi}_{0X}^{vir}}_{A_1BE} + \sqrt{\frac{1 - \braket{ \tilde{\psi}_{0Z} | \tilde{\psi}_{1Z}}_{A_1BE}}{2}} \ket{1_X}_{A_2} \ket{\tilde{\psi}_{1X}^{vir}}_{A_1BE},
\end{equation}
where
\begin{equation}
\ket{\tilde{\psi}_{jX}^{vir}}_{A_1BE} = \frac{ \ket{\tilde{\psi}_{0Z}}_{A_1BE} + (-1)^j  \ket{\tilde{\psi}_{1Z}}_{A_1BE}}{\sqrt{2 \big(1 + (-1)^j \braket{ \tilde{\psi}_{0Z} | \tilde{\psi}_{1Z}}_{A_1BE}\big)}}.
\end{equation}

In the virtual protocol, we consider that Alice sends Bob two virtual states, $\ket{\tilde{\psi}_{jX}^{vir}}_{A_1BE}$, and three actual states, $\ket{\tilde{\psi}_{jZ}}_{A_1BE}$ and $\ket{\tilde{\psi}_{0X}}_{A_1BE}$, which are used to estimate the phase error rate. We have seen that even in the case of mixed states we can define actual and virtual pure states, and these pure states can be directly used in our security proof. Therefore, our formalism is valid for mixed states in a single-mode qubit space.

\begin{figure}[H]
	\centering
	\includegraphics[width=11cm]{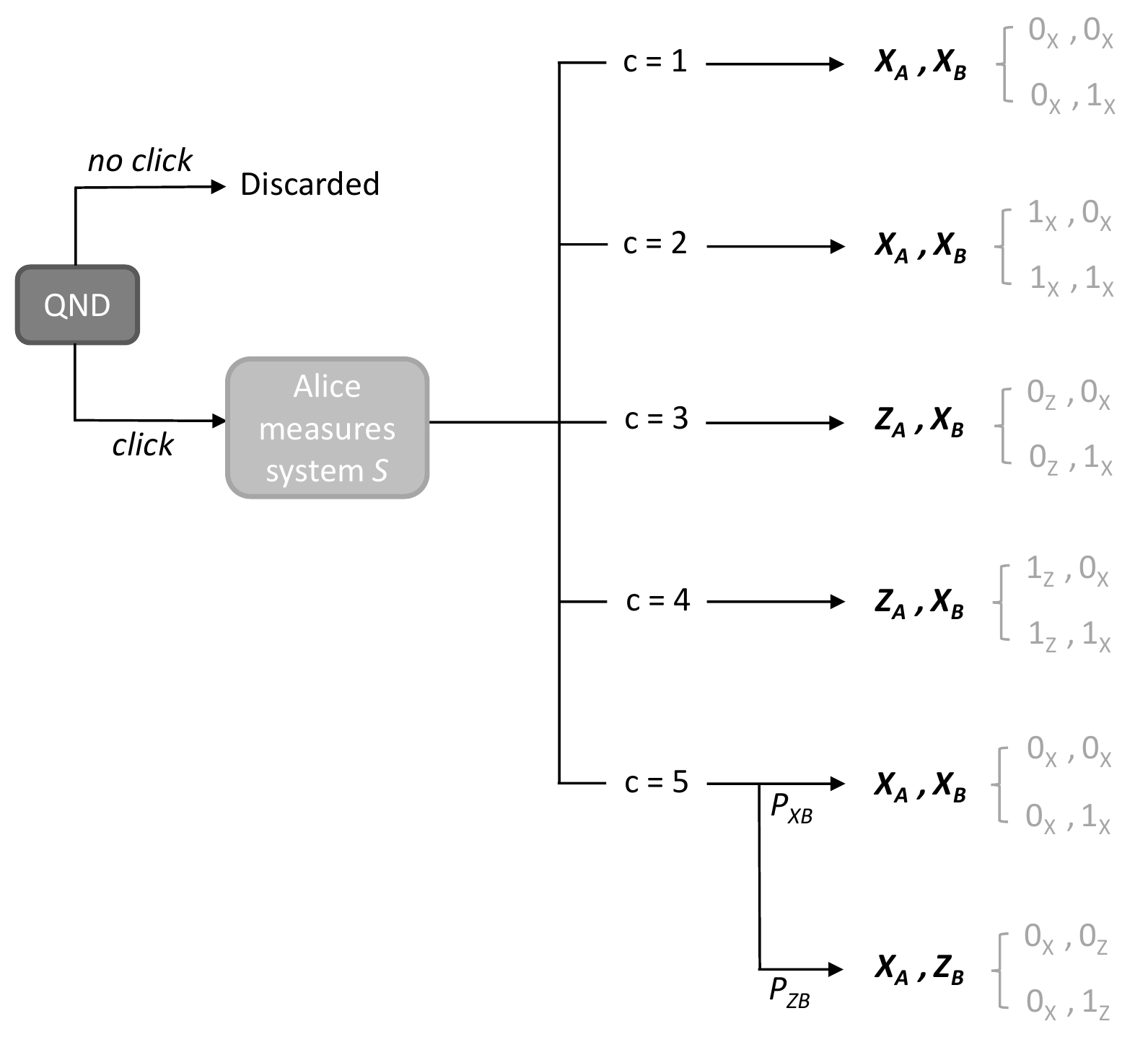} 
	\caption{The logical schematics for the virtual protocol, where the notation $X_A/Z_A$, $X_B/Z_B$ corresponds to Alice's and Bob's measurements bases respectively. The virtual states correspond to $c = 1,2$, the actual $Z$ states to $c = 3,4$, and the actual $X$ states to $c =5$. For each click event, Alice measures system $S$ and Bob measures system $BE$. Note that, the selection of $c = 1,2,3,4$ already includes Bob's measurement in the $X$ basis, but when $c = 5$ his measurement basis is chosen probabilistically.}
	\label{fig:virtualLT}
\end{figure}

Next, let us continue to explain the security proof in more detail. The selection of these actual and virtual states can be expressed as
\begin{equation}
\ket{\varphi}_{SA_1BE} = \sum_{c=1}^5  \sqrt{P(c)} \ket{c}_{S}  \ket{\vartheta^{(c)}}_{A_1BE},
\label{eq:prep}
\end{equation}
where $S$ is the shield system that is kept inside of Alice's lab and the states $\ket{\vartheta^{(c)}}_{A_1BE}$ are 
\begin{equation}
\begin{split}
&\ket{\vartheta^{(1)}}_{A_1BE} = \ket{\tilde{\psi}_{0X}^{vir}}_{A_1BE}, \\
&\ket{\vartheta^{(2)}}_{A_1BE} = \ket{\tilde{\psi}_{1X}^{vir}}_{A_1BE}, \\
&\ket{\vartheta^{(3)}}_{A_1BE} = \ket{\tilde{\psi}_{0Z}}_{A_1BE}, \\
&\ket{\vartheta^{(4)}}_{A_1BE} = \ket{\tilde{\psi}_{1Z}}_{A_1BE}, \\
&\ket{\vartheta^{(5)}}_{A_1BE} = \ket{\tilde{\psi}_{0X}}_{A_1BE}, \\
\end{split}
\end{equation}
with their respective probabilities $P(c)$
\begin{equation}
\begin{split}
&P(1) = \frac{{P_{Z_A}}{P_{Z_B}}}{2} \Big(1 + \braket{ \tilde{\psi}_{0Z} | \tilde{\psi}_{1Z}}_{A_1BE} \Big), \\
&P(2) = \frac{{P_{Z_A}}{P_{Z_B}}}{2} \Big(1 - \braket{ \tilde{\psi}_{0Z} | \tilde{\psi}_{1Z}}_{A_1BE} \Big), \\
&P(3) = \frac{{P_{Z_A}}{P_{X_B}}}{2}, \\
&P(4) = \frac{{P_{Z_A}}{P_{X_B}}}{2}, \\
&P(5) = P_{X_A}{P_{Z_B}} + P_{X_A}{P_{X_B}} = P_{X_A}. \\
\end{split}
\end{equation}
When Bob receives the states he performs a measurement in either the $Z$ or the $X$ basis, and these are defined by the POVMs described in Section \ref{sec:description}. Also, all announcements between Alice and Bob are done via an authenticated public channel. Note that, in the virtual protocol we assume that Alice and Bob are sitting in the same lab so that they can choose the measurement basis, and this is allowed because the quantum and classical information available to Eve is the same between the actual and the virtual protocols. The detailed steps of the virtual protocol are presented below and the logic schematics in Fig. \ref{fig:virtualLT}. \\

\begin{enumerate}
\item \textbf{Initialisation:} Before running the protocol, Alice and Bob agree on a number $N_{fixed}$ of rounds, on the error correcting codes, and on a set of hash functions to perform privacy amplification. Steps 2-4 of the protocol are repeated $N$ times until the number of detected events $N$ becomes $N_{fixed}$. 

\item \textbf{State Preparation:} After a potential THA, Alice prepares systems $S$, $A_1$ and $BE$ in the entangled state $\ket{\varphi}_{SA_1BE}$, in Eq. (\ref{eq:prep}), and sends Bob the system $BE$ via a quantum channel. 

\item \textbf{QND measurement:} For each incoming system, Bob performs a quantum non-demolition (QND) measurement to determine whether the signals are detected or not. If Bob obtains a detection event he keeps the resulting system and $N$ is increased by 1 unit. 

\item \textbf{Detection announcement:} If $N=N_{fixed}$, Bob announces the termination of quantum communication and the detection pattern. Otherwise Alice and Bob return to Step 2 of the protocol. 

\item \textbf{Measurement and basis announcement:} For each of the detected events, Alice measures her system $S$ and announces the $Z$ $(X)$ basis when $c = 1,2,3,4 ~ (c = 5)$. Bob announces the $Z$ $(X)$ basis for $c = 1,2 ~(3,4)$, but he always measures in the $X$ basis. For $c =5$, Bob selects the basis $\beta \in\{ Z, X\}$ probabilistically and announces his basis choice. Then, he carries out the measurement on system $BE$ in his selected basis.

\item \textbf{Sifting and announcement:} Alice and Bob define and announce the bit strings $\vec{s}_{X,0Z}$, $\vec{s}_{X,1Z}$ and $\vec{s}_{X,0X}$, which correspond to the events when Alice sends the actual states and Bob performs the $X$ basis measurements. These are the basis mismatched events when $c = 3,4$ and one of the events when $c = 5$, the basis matched event. These strings are used to estimate the phase error rate. 

\end{enumerate} 

In the virtual protocol, we require that Alice and Bob postpone their measurements until the quantum communication ends, therefore, we assume that Alice and Bob possess quantum memories where they can store their systems. The reason for this deferral comes from the application of Azuma's inequality, which is explained later. In the case of Alice, she only makes her measurement after the termination condition, in Step 5. This is allowed because it does not matter when she performs the measurement since it commutes with Eve's operations and hence it will not affect Alice's statistics. For Bob, we divide his measurement in two steps: a QND measurement, which allows him to know when a detected event occurred, and a measurement to output the bit value with the chosen basis. If the QND measurement results in a detected instance, Bob performs the measurement using the $Z$ or $X$ basis. We are able to delay Bob's measurement choice because the inconclusive outcomes are assumed to be independent of the basis, as explained in Section \ref{sec:description}. The key point in the virtual protocol is as follows: the security of the events when Alice sends the actual $Z$ basis states and Bob obtains a detected event in the actual protocol with the $Z$ basis, can be analysed by imagining that Alice and Bob both employ the $X$ basis to measure respectively the systems $A_2$ and $BE$. This means that when Alice sends a virtual state $(c=1,2)$ Bob's measurement basis is always the $X$ basis. 

It is clear that the virtual protocol described here is equivalent to the actual protocol in Section \ref{sec:description}. This is so because the quantum states sent by Alice are the same in both protocols as well as the announcements made by the two parties. For instance, when Alice sends the virtual states they both measure in the $X$ basis but they announce the $Z$ basis (Step 5). In the actual protocol, these events are used for key generation and therefore Alice and Bob also announce the $Z$ basis. This means that the protocols are indistinguishable from Eve's perspective as required. Note that, the virtual protocol does not produce a key, it is merely used for the estimation of the phase error rate.

\subsection{Azuma's inequality and its application to the security proof}
In coherent attacks, Eve interacts with all the signals sent by Alice followed by a joint measurement after listening to all the classical information exchanged between Alice and Bob. In this scenario we use Azuma's inequality \cite{azuma} which takes into account this dependency, and allows us to derive a relation between the expected values and the observed values. Most importantly, once we have the conditional probabilities on all previous measurement outcomes we can to find the actual number of events observed.

Azumas's inequality can be applied to a stochastic model as long as a sequence of random variables is a martingale and satisfies the bounded difference conditions (BDC). A Martingale is a sequence of random variables $X^{(0)}, X^{(1)},..., X^{(l)}$ for which the expectation $E[\cdot]$ of the next value is equal to the present value in the sequence given that we know all the previous outcomes, i.e., $E [X^{(l+1)} | X^{(0)}, X^{(1)},..., X^{(l)}] = X^{(l)}$ for all $l \ge 0$. This sequence is said to satisfy BDC if there exists $c^{(l)} > 0$ such that $|X^{(l+1)} - X^{(l)}| \le c^{(l)}$ for all  $l \ge 0$. For $N$ trials of a variable $X^{(l)}$ with $c^{(l)} = 1$, Azuma's inequality states that
\begin{equation}
P \big[|X^{(N)} - X^{(0)}| > N \delta_A \big] \le 2e^{\frac{-N\delta_A^2}{2}},
\end{equation}
holds for any $\delta_A \in (0,1)$. Now, for the $l$th trial, we define $X^{(l)}$ as 
\begin{equation}
X^{(l)} := \Lambda^{(l)} - \sum_{k=1}^{l} P(\zeta_k = 1| \zeta_0, ..., \zeta_{k-1}),
\label{eq:variable}
\end{equation}
where $\Lambda^{(l)}$ is a random variable representing the actual number of events (that is $\Lambda^{(l)} = \sum_{k=1}^l \zeta_k$ ) observed during the first $l$ trials, $\zeta_k$ is the random variable of interest and it has the value of 0 or 1. Moreover, $P(\zeta_k = 1| \zeta_0, ..., \zeta_{k-1})$ is the conditional probability of obtaining the outcome specified by $\zeta_k = 1$ in the $k$th trial given that the first $k-1$ outcomes are $ \zeta_0, ..., \zeta_{k-1}$. It is possible to show that the sequence of random variables in Eq. (\ref{eq:variable}) is Martingale and satisfies the BDC. Hence, we can apply the Azuma's inequality and write
\begin{equation}
P \Bigg[\bigg|\Lambda^{(N)} - \sum_{k=1}^{N} P(\zeta_k = 1| \zeta_0, ..., \zeta_{k-1}) \bigg| > N \delta_A \Bigg] \le 2e^{\frac{-N\delta_A^2}{2}},
\end{equation}
where we use the definition $X^{(0)} = 0$. This also means that 
\begin{equation}
\sum_{k=1}^{N} P(\zeta_k = 1| \zeta_0, ..., \zeta_{k-1}) - N \delta_A \le \Lambda^{(N)} \le  \sum_{k=1}^{N} P(\zeta_k = 1| \zeta_0, ..., \zeta_{k-1}) + N \delta_A,
\end{equation}
holds at least with probability $P = 1- 2e^{\frac{-N\delta_A^2}{2}}$. Therefore,
\begin{equation}
\Lambda^{(N)} = \sum_{k=1}^{N}  P(\zeta_k = 1| \zeta_0, ..., \zeta_{k-1}) + \delta_B,
\label{eq:lambda}
\end{equation} 
except for error probability $\epsilon + \hat{\epsilon}$, where the deviation parameter $\delta_B \in [-\Delta,\hat{\Delta}]$. These bounds are defined as $\Delta = f(N,\epsilon)$ and  $\hat{\Delta} = f(N,\hat{\epsilon})$ where $f(x,y) = \sqrt{2x \ln(1/y)}$.

Let us now show how we use this inequality in our security proof. In particular, we consider 
%
\begin{equation}
X^{(l)}_{csX} = \Lambda^{(l)}_{csX} - \sum_{k=1}^{l} P(\zeta_{k,csX} = 1| \zeta_{0}, ..., \zeta_{k-1}),
\label{eq:variable2}
\end{equation}
where $csX = c,s$ for $c = 1,2,3,4$ since Bob's basis choice is already included in these cases, and $csX = c,s,X$ for $c = 5$. In Eq. (\ref{eq:variable2}), $P(\zeta_{k,csX} = 1| \zeta_{0}, ..., \zeta_{k-1})$ is the probability of Alice selecting the state $c$ and Bob observing $s$ ($s,X$) for $s \in \{0,1\}$ when $c = 1,2,3,4$ ($c=5$) in the $k$th trial, conditional on all the previous outcomes from the measurements $ \zeta_0, ..., \zeta_{k-1}$. To obtain this probability we first define
\begin{equation}
\ket{\tau}_{SA_1BE} = \ket{\varphi_{k-1}}_{SA_1BE}\ket{\varphi_k}_{SA_1BE}\ket{\varphi_r}_{SA_1BE},
\end{equation}
to be the state prepared by Alice in an execution of the protocol, where $\ket{\varphi_{k-1}}_{SA_1BE}$, $\ket{\varphi_k}_{SA_1BE}$ and $\ket{\varphi_r}_{SA_1BE}$ correspond to all the systems before the $k$th trial, in the $k$th trial and in the rest of the trials after $k$ (i.e., $r = N-k$), respectively. 

Eve's action can be described as $\hat{U}_{BEE'} \ket{\tau}_{SA_1BE} \ket{0}_{E'} = \sum_t \hat{B}_{tB} \ket{\tau}_{SA_1BE} \ket{t}_{E'}$, where $\hat{U}_{BEE'}$ is a unitary transformation acting on systems $BEE'$, $\hat{B}_{tB}$ is the Kraus operator which acts on system $BE$ depending on Eve's measurement outcome $t$, and $\ket{t}_{\{t= 1,2,...\}}$ is an orthonormal basis. Note that, here we use the subscript $E$ to refer to Eve's system originating from a $THA$ and $E'$ corresponds to the additional ancilla system in her hands. Alice and Bob only communicate after performing the measurements so these parameters are independent of the state preparation. 

In order to consider Alice's and Bob's measurements previous to the $k$th trial, we define the operator $\hat{\mathcal O}_{k-1,SBE} = \otimes^{k-1}_{\nu=1} \hat{M}_{S_{\nu}{BE}_{\nu}}$, where $\hat{M}_{S_{\nu}{BE}_{\nu}}$ denotes the Kraus operator associated with the $\nu$th measurement outcome of Alice and Bob. Hence, after Eve's interaction, the normalised $k$th state of the system $SBE$ conditioned on the measurement outcomes, $O_{k-1}$, and the detected event can be expressed as 
\begin{equation}
\hat{\rho}_{k|O_{k-1}}^{SBE} = \frac{\hat{\sigma}_{k|O_{k-1}}^{SBE}}{\Tr (\hat{\sigma}_{k|O_{k-1}}^{SBE})},
\end{equation} 
where the state $\hat{\sigma}_{k|O_{k-1}}^{SBE}$ is defined shortly below (see Eq. (\ref{eq:trace})). We know that 
\begin{equation}
\hat{\sigma}_{k|O_{k-1}}^{SA_1BE}= \sum_t \Tr_{\bar{k}} \Big[ \hat{F}_{BE_k} \hat{\mathcal O}_{k-1,SBE} \hat{B}_{tB} \ket{\tau}_{SA_1BE } \bra{\tau} \hat{B}_{tB}^\dagger \hat{\mathcal O}_{k-1,SBE}^\dagger \hat{F}_{BE_k}^\dagger \Big],
\label{eq:sigma}
\end{equation}
where $\Tr_{\bar{k}}$ is the partial trace over the systems $S$, $A_1$ and $BE$ for all the events that are not in the $k$th trial, and $\hat{F}_{BE_k}$ is Bob's Kraus operator acting on the $k$th system, corresponding to the detected events. This means taking the trace with the basis $\{ \ket{\vec{x}_{k-1}} , \ket{\vec{x}_r}\}$, where $\ket{\vec{x}_{k-1}}$ corresponds to all the systems in the first $k-1$ runs and $\ket{\vec{x}_r}$ to the rest of the systems after $k$. Then, we can rewrite Eq. (\ref{eq:sigma}) as
\begin{equation}
\hat{\sigma}_{k|O_{k-1}}^{SBE} =  \sum_t \sum_{\vec{x}_{k-1}, \vec{x}_r} \Tr_{A_1}^k \Big[A_{t,BE|O_{k-1}}^{(\vec{x}_{k-1}, \vec{x}_r)} \ket{\varphi_k}_{SA_1BE} \bra{\varphi_k} A_{t,BE|O_{k-1}}^{\dagger(\vec{x}_{k-1}, \vec{x}_r)}\Big],
\label{eq:trace}
\end{equation}
where $\Tr_{A_1}^k$ is the partial trace over the system $A_1$ in the $k$th trial and $A_{t,BE|O_{k-1}}^{(\vec{x}_{k-1}, \vec{x}_r)}$ is the Kraus operator acting on the $k$th system conditional on all the previous detected events, and it is defined as
\begin{equation}
A_{t,BE|O_{k-1}}^{(\vec{x}_{k-1}, \vec{x}_r)}  = \bra{\vec{x}_r} \bra{\vec{x}_{k-1}} \hat{F}_{BE_k} \hat{O}_{k-1,SBE} ~\hat{B}_{tB}  \ket{\varphi_{k-1}}_{SA_1BE} \ket{\varphi_r}_{SA_1BE}.
\end{equation}

By substituting Eq.(\ref{eq:prep}) into Eq.({\ref{eq:trace}}) we get 
\begin{equation}
\hat{\sigma}_{k|O_{k-1}}^{SBE}  = \sum_{c,c'} \sqrt{P(c) P(c')} ~\sum_t \sum_{\vec{x}_{k-1}, \vec{x}_r} \Tr_{A_1}^k \Big[A_{t,BE|O_{k-1}}^{(\vec{x}_{k-1}, \vec{x}_r)}  \ket{c}_S\bra{c'} \otimes \ket{\vartheta^{(c)}}_{A_1BE} \bra{\vartheta^{(c')}}  A_{t,BE|O_{k-1}}^{\dagger(\vec{x}_{k-1}, \vec{x}_r)}\Big]. 
\end{equation}
It is clear now that this state is dependent on Eve's action as well as on the previous outcomes. Also, note that the partial trace only acts on system $A_1$. The probability that Alice obtains the outcome $c$, Bob selects the $X$ basis and obtains a bit value $s$ conditional on all the previous measurement outcomes is calculated as 
\begin{equation}
\begin{split}
P_{csX|O_{k-1}}  &= \frac{P(X \cap c)}{ \Tr(\hat{\sigma}_{k|O_{k-1}}^{SBE})} \sum_t \sum_{\vec{x}_{k-1}, \vec{x}_r} \Tr \Bigg[A_{t,BE|O_{k-1}}^{(\vec{x}_{k-1}, \vec{x}_r)}  \Tr_{A_1}^k \Big[ \dyad*{\vartheta^{(c)}}{\vartheta^{(c)}}_{A_1BE} \Big]  A_{t,BE|O_{k-1}}^{\dagger(\vec{x}_{k-1}, \vec{x}_r)} \hat{M}_{sX}\Bigg] \\
&=  \frac{P(X \cap c)}{\Tr(\hat{\sigma}_{k|O_{k-1}}^{SBE})} \Tr \Bigg[ \hat{D}_{sX|O_{k-1}}  \Tr_{A_1}^k \Big[ \dyad*{\vartheta^{(c)}}{\vartheta^{(c)}}_{A_1BE}\Big] \Bigg] \\
\end{split}
\end{equation}
where $P(X \cap c) = P(c)$ for $c = 1,2,3,4$ and $P(X \cap c) = P(c)P(X_B)$ for $c =5$. In this expression, $\hat{D}_{sX|O_{k-1}} =  \sum_t \sum_{\vec{x}_{k-1}, \vec{x}_r} A_{t,BE|O_{k-1}}^{\dagger(\vec{x}_{k-1}, \vec{x}_r)}  \hat{M}_{sX}  A_{t,BE|O_{k-1}}^{(\vec{x}_{k-1}, \vec{x}_r)} $ represents Eve's action as well as Bob's measurement. This is independent of $c$, which means that Eve cannot behave differently depending of the state sent. Importantly, the probability $P_{csX|O_{k-1}}$ essentially corresponds to the actual yields $Y_{sX,j\beta}$ in the main text when $c = 3,4,5$. Note that the yields in this Appendix are normalised by the detected events while those in the main text are not. In the finite key size regime, the normalisation according to the detected events results in a better performance, however, in the limit of large number of pulses, they are essentially the same. As we consider this limit throughout this paper, in the main text we adopt the yields that are not normalised by the detected events for simplicity of explanation. We know $\Lambda_{3sX}$,  $\Lambda_{4sX}$ and $\Lambda_{5sX}$ by collecting the corresponding number of events from the actual protocol. Therefore using Azuma's inequality, i.e., Eq. (\ref{eq:lambda}), we can calculate the conditional probabilities which correspond to the yields $Y_{sX,0Z}^{(Z)}$, $Y_{sX,1Z}^{(Z)}$ and $Y_{sX,0X}^{(X)}$ respectively. From Section \ref{sec:estimation}, we know how these yields are related to the transmission rates, and in turn, how these are related to the virtual yields $Y_{1X,0Z}^{(Z)vir}$ and $Y_{0X,1Z}^{(Z)vir}$. Here, we would like to emphasise that using Eq. (\ref{eq:coef2}) we can calculate these yields, which correspond to the probabilities $P_{11X|O_{k-1}}$ and $P_{20X|O_{k-1}}$ respectively, both of which are conditional on the previous measurement outcomes. Using Azuma's inequality again, we can find the number of number of events, $\Lambda_{11X}$ and $\Lambda_{20X}$, which are the number of phase errors, and this concludes the estimation of the phase error rate.  

\section{Coefficients}
\label{app:coefficients}
In this appendix, we list the coefficients used in the main text. Direct calculations show that the coefficients $A_j$, $B_j$ and $C_j$ for Eqs. (\ref{eq:yield1})-(\ref{eq:coef2}) are given by 
\begin{linenomath*}\begin{equation}
\begin{split}
A_j = & \frac{1}{4} \Big[|a_{0Z}|^2 + (-1)^j \Big( a_{0Z}^* a_{1Z}{\braket{\phi_{0Z}|\phi_{1Z}}}_{BE} + a_{0Z} a_{1Z}^* {\braket{\phi_{1Z}|\phi_{0Z}}}_{BE}  \Big) + |a_{1Z}|^2 \Big],\\ 
B_j = & \frac{1}{4} \sqrt{|a_{0Z}|^2 + (-1)^j \Big(a_{0Z}^* a_{1Z} {\braket{\phi_{0Z}|\phi_{1Z}}}_{BE} + a_{0Z} a_{1Z}^* {\braket{\phi_{1Z}|\phi_{0Z}}}_{BE}  \Big) + |a_{1Z}|^2}\\
& \times ~\sqrt{|b_{0Z}|^2 + (-1)^j \Big(b_{0Z}^* b_{1Z} {\braket{\phi_{0Z}^\perp|\phi_{1Z}^\perp}}_{BE} + b_{0Z} b_{1Z}^2 {\braket{\phi_{1Z}^\perp|\phi_{0Z}^\perp}}_{BE}  \Big) + |b_{1Z}|^2},\\
C_j= &\frac{1}{4} \Big[|b_{0Z}|^2 + (-1)^j \Big( b_{0Z}^* b_{1Z}{\braket{\phi_{0Z}^\perp|\phi_{1Z}^\perp}}_{BE} + b_{0Z} b_{1Z}^*{\braket{\phi_{1Z}^\perp|\phi_{0Z}^\perp}}_{BE}  \Big) + b_{1Z}^2 \Big].\\
\end{split}
\end{equation}\end{linenomath*} 

\section{Lo-Preskill's analysis}
\label{app:lopreskill}
In this section, we outline the security argument of Lo-Preskill's analysis \cite{lo4} and apply it to the particular device model described in the main text.

\subsection{Description of Lo-Preskill's protocol}
In order to prove the security of the BB84 protocol \citep{bennett}, where the state of the emitted pulses is not phase randomised, we convert the actual protocol into an entanglement based protocol. In this protocol, the $Z$ basis states are prepared when Alice measures her system $A$ of the entangled state 
\begin{equation}
\ket{\Upsilon_Z}_{ABE} = \frac{1}{\sqrt{2}} \big[\ket{0_Z}_A \otimes \ket{\Phi_{0Z}}_{BE} + \ket{1_Z}_A \otimes \ket{\Phi_{1Z}}_{BE} \big],
\label{eq:ent1}
\end{equation}
with the basis $\{ \ket{0_Z}, \ket{1_Z}\}$. Similarly, she prepares the $X$ basis states by measuring her system $A$ of the following entangled state in the basis $\{ \ket{0_X}, \ket{1_X}\}$
\begin{equation}
\ket{\Upsilon_X}_{ABE} = \frac{1}{\sqrt{2}} \big[\ket{0_X}_A\otimes \ket{\Phi_{0X}}_{BE} + \ket{1_X}_A \otimes \ket{\Phi_{1X}}_{BE} \big].
\label{eq:ent2}
\end{equation}

These states contain Eve's system $E$ since we are considering a potential THA, and the states $\ket{\Phi_{j\beta}}$ for $j \in \{0,1\}$ and $\beta \in \{Z,X\}$ are defined in Eq. (\ref{eq:multi-mode}). If the state preparation is perfect and the source is completely isolated there is no passive or active leakage of information, which is the case when $\delta = \theta_{j\beta} = \mu = 0$. This means that these states are basis independent, i.e., $\ket{\Upsilon_Z}_{ABE} = \ket{\Upsilon_X}_{ABE}$. When the state preparation is not perfect (at least one of the quantities $\delta$, $\theta_{j\beta}$, $\mu$ is greater than zero), the states $\ket{\Upsilon_Z}_{ABE}$ and $\ket{\Upsilon_X}_{ABE}$ are close to each other but not equal hence, some information related with the basis choice could be leaked from the source. This means that Eve might be able to distinguish the states emitted when the $Z$ basis is chosen from the ones when the $X$ basis is chosen. This basis-dependency can be quantified by considering an equivalent virtual protocol that evaluates the ``balance" of a ``quantum coin" \cite{gottesman}. To investigate this, we consider that Alice prepares the state 
\begin{equation}
\ket{\Gamma}_{CABE} = \frac{1}{\sqrt{2}} \big[\ket{0_Z}_C \otimes \ket{\Upsilon_Z}_{ABE} + \ket{1_Z}_C \otimes \ket{\Upsilon_X}_{ABE} \big],
\label{eq:phistate}
\end{equation}
where $C$ corresponds to the quantum coin. She performs a measurement on the coin with the basis $\{ \ket{0_Z}, \ket{1_Z}\}$ to determine the encoding of the signal. From Koashi's approach \cite{koashi2,koashi3}, we can find the phase errors using the complementary scenario, by applying the ``Bloch sphere bound" \cite{tamaki2} to the quantum coin. For this, we consider switching the measurement basis (to measure with the $X$ basis instead of the $Z$ basis) on the quantum coin, thus it is useful to write Eq.(\ref{eq:phistate}) as 
\begin{equation}
\ket{\Gamma}_{CABE}  = \frac{1}{2} \big[ \ket{0_X}_C \big(\ket{\Upsilon_Z}_{ABE} + \ket{\Upsilon_X}_{ABE} \big) + \ket{1_X}_C \big(\ket{\Upsilon_Z}_{ABE} - \ket{\Upsilon_X}_{ABE} \big)\big].
\label{eq:imbalance}
\end{equation}

For characterising how close these states are, we employ the probability $\Delta$ associated with finding the state $\ket{1_X}_C$ and the outcome $X = 1$.  This can be expressed as
\begin{equation}
\Delta = \frac{1}{4} \big(2 - \braket{\Upsilon_Z | \Upsilon_X}_{ABE} - \braket{\Upsilon_Z | \Upsilon_X}^*_{ABE} \big),
\end{equation}
where $\braket{\Upsilon_Z | \Upsilon_X}^*_{ABE}$ is the complex conjugate of $\braket{\Upsilon_Z | \Upsilon_X}_{ABE}$. These overlaps can be calculated to obtain $\Delta$, which quantifies the basis dependence of Alice's states. If there are no imperfections, the inner product $\braket{\Upsilon_Z | \Upsilon_X}_{ABE} = \braket{\Upsilon_Z | \Upsilon_X}^*_{ABE} = 1$ and therefore the coin imbalance is $\Delta = 0$. Note that, this justifies our choices of states in Eqs. (\ref{eq:ent1}) and (\ref{eq:ent2}). Similarly, when they are completely orthogonal to each other the inner products are zero and $\Delta = \frac{1}{2}$. This means that, the closer these states are to each other, the smaller is the coin imbalance $\Delta$ and therefore, less information is leaked to Eve.

In a QKD scheme, not all the signals sent from the source are detected due to channel loss. Therefore, Eve might take advantage of this flaw by blocking certain signals that are not favourable to her without being detected, enhancing the basis dependence of the signals. To take this into consideration, the Lo-Preskill's analysis assumes the worst case scenario, where it maximises the imbalance of the coin, or in other words, maximises the leakage of information to Eve. This means that it assumes that all the signals that do not produce a click on Bob's side are associated with the outcome $X=0$ when measuring the quantum coin. The accommodation of this worst case scenario is reflected by considering an enhancement probability of $\Delta$ such that: $\Delta' = \frac{\Delta}{min[Y_Z, Y_X]}$ where $Y_Z$ $(Y_X)$ is the single photon yield in the $Z$ $(X)$ basis \cite{lucamarini}.

In order to find the secret key rate $R$ we need to estimate the phase error rate $e_X$, which corresponds to what the bit error rate would have been if Alice and Bob had measured their states in the $X$ basis when the entangled state prepared is $\ket{\Upsilon_Z}_{ABE}$. In the Lo-Preskill's analysis, $e_X$ cannot be calculated directly from the channel model but using $e_Z$ and the coin imbalance $\Delta'$ the phase error rate can be estimated to be \cite{lo4}
\begin{equation}
e_X  \le {e_Z} + 4 \Delta' (1-\Delta') (1 - 2 {e_Z}) + 4 (1 - 2\Delta' ) \sqrt{\Delta' (1 - \Delta') {e_Z} (1 - {e_Z})}.
\label{eq:eph_delta}
\end{equation}
In the ideal scenario, when the states $\ket{\Upsilon_Z}_{ABE}$ and $\ket{\Upsilon_X}_{ABE}$ are close to each other, $\Delta'$ is very small and $e_X \approx e_Z$. By substituting Eq. (\ref{eq:eph_delta}) into Eq. (\ref{eq:keyrate}) and using the same definition for $e_Z$ as in Eq. (\ref{eq:biterror}) we are able to calculate $R$. 

\subsection{Actual and virtual protocols}
In this section, we describe the actual and virtual QKD protocols considered in the Lo-Preskill's security analysis. Note that, all announcements between Alice and Bob are done via an authenticated public channel. \\

\textbf{Actual Protocol} 
\begin{enumerate}
\item \textbf{Initialisation:} Before running the protocol, Alice and Bob agree on a number $N_{fixed}$ of rounds, on the error correcting codes, and on a set of hash functions to perform privacy amplification. Steps 2-4 of the protocol are repeated $N$ times until the number of detected events $N$ becomes $N_{fixed}$.  

\item \textbf{State preparation:} Alice selects the basis $\beta \in\{ Z, X\}$ for encoding the states with probabilities $P_{Z_A}$ and $P_{X_A} = 1- P_{Z_A}$ respectively. For each basis selected, she randomly chooses a bit value and the associated phase. Then, she prepares the signal and reference pulses following these specifications and sends the state to Bob via the quantum channel. Due to a potential THA, the sent states might contain Eve's system. 

\item \textbf{Measurement:} Bob measures each incoming signal using the basis $\beta \in \{Z,X\}$ with probabilities $P_{Z_B}$ and $P_{X_B} = 1 - P_{Z_B}$ respectively. For each detected event, $N$ is increased by 1 unit.  

\item \textbf{Detection announcement:} If $N=N_{fixed}$, Bob announces the termination of the quantum communication phase and the detection pattern. Otherwise, Alice and Bob return to Step 2 of the protocol. 

\item \textbf{Basis announcement and sifting:} For the detected events, Alice selects the basis of the quantum coin namely, the $Z_C$ or $X_C$ basis, with probabilities $P_{Z_{C}}$ and $P_{X_{C}}= 1- P_{Z_{C}}$ respectively. She announces the chosen basis, and if $Z_C$ was selected, she also announces the selected basis in Step 2. Moreover, Bob announces his basis choice. If Alice's and Bob's basis choices disagree they discard the data. Otherwise, they define bit strings associated with the matched events, i.e., when both select the $Z$ or both select the $X$ basis. 

\item \textbf{Parameter estimation:} Alice and Bob announce the bit strings in the $X$ basis. They calculate the Hamming weight \texttt{wt}$(\vec{s}_{X_A} \oplus \vec{s}_{X_B})$ to find the number of mismatches between the two bit strings. This is then used to estimate the number of bits that need to be removed from the sifted strings in the $Z$ basis, $\vec{s}_{Z_A}$ and $\vec{s}_{Z_A}$, during privacy amplification. 

\item \textbf{Error correction and privacy amplification:} Alice and Bob randomly select an error correcting code from Step 1 to perform error correction on the sifted strings in the $Z$ basis and then they exchange the syndrome information. Then, by choosing a random hash function from Step 1 and based on the result of the parameter estimation in Step 6, they perform privacy amplification on the corrected sifted keys. At the end, Alice and Bob obtain the key $\vec{k}_{Z_A}$ and $\vec{k}_{Z_B}$, respectively. 

\end{enumerate} 

\vspace{10pt}

\textbf{Virtual Protocol} \\

\begin{figure}[H]
	\centering
	\includegraphics[width=17cm]{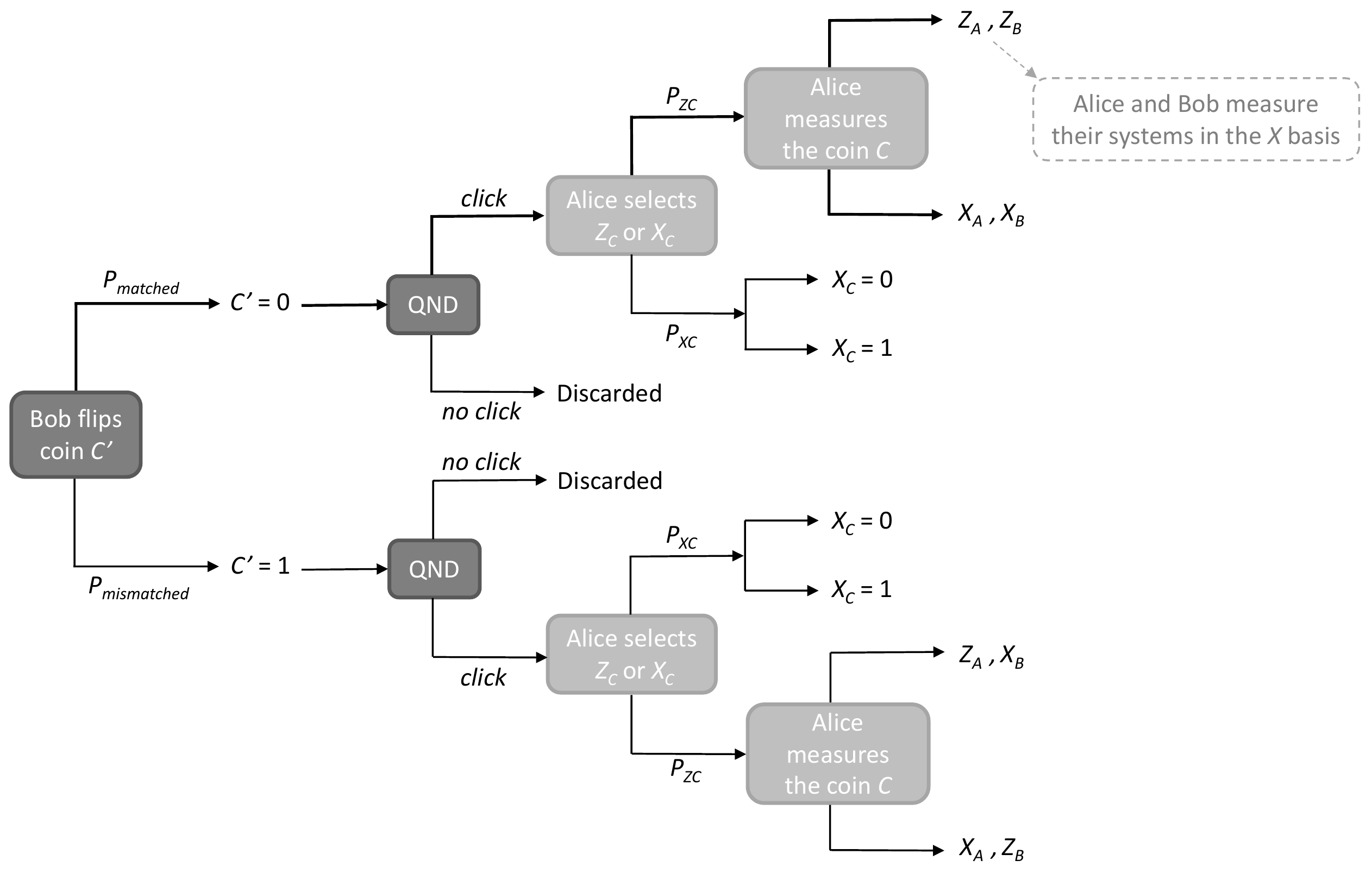} 
	\caption{The logical schematics for the virtual protocol. It is the use of a second coin that allows us to pre-select only the matched events. When Alice selects the $Z$ basis to measure the coin $C$ she prepares the state $\ket{\Upsilon_Z}$ or $\ket{\Upsilon_X}$, but in the virtual protocol she always measures in the $X$ basis. }
	\label{fig:virtualLP}
\end{figure}

In Lo-Preskill's work, they consider only basis matched events therefore, in the virtual protocol we need to include a post-selection that will only choose these events. This can be done by flipping a classical coin $C'$ that selects between the basis matched and mismatched events. The probability that $C'=0$ $(1)$ is $P_{matched}$ $(P_{mismatched})$, associated with the basis matched (mismatched) event chosen by Bob. A diagram of the virtual protocol with this post-selection is depicted in Fig. \ref{fig:virtualLP}. Also, all announcements between Alice and Bob are done via an authenticated public channel. The steps of the protocol are as follows:  

\begin{enumerate}
\item \textbf{Initialisation:} Before running the protocol, Alice and Bob agree on a number $N_{fixed}$ of rounds, on the error correcting codes, and on a set of hash functions to perform privacy amplification. Steps 2-5 of the protocol are repeated until the number of detected events $N$ becomes $N_{fixed}$. 

\item \textbf{Probabilistic choice of the basis matched and mismatched events:} Bob measures the classical coin $C'$ and obtains the results ``0" or ``1", which correspond to basis matched and mismatched events, respectively. 

\item \textbf{State Preparation:} After a potential THA, Alice prepares systems in the entangled state in Eq. (\ref{eq:phistate}) and sends Bob the system $BE$ via a quantum channel. 

\item \textbf{QND measurement:} For each incoming system, Bob performs a quantum non-demolition (QND) measurement to determine whether the signal is detected or not. If Bob obtains a detection event he keeps the resulting system and $N$ is increased by 1 unit. 

\item \textbf{Detection announcement:} If $N=N_{fixed}$, Bob announces the termination of the quantum communication phase and the detection pattern. Otherwise Alice and Bob return to Step 2 of the protocol. 

\item \textbf{Measurement and basis announcement:} For the detected events, Alice measures the quantum coin with the $Z_C$ basis or the $X_C$ basis, chosen with probabilities $P_{Z_{C}}$ and $P_{X_{C}} = 1- P_{Z_{C}}$ respectively. Then, she announces the chosen basis. If $Z_C$ is selected she also announces the measurement outcome (the $Z$ or the $X$ basis), however, she always measures system $A$ with the $X$ basis. As for Bob, if $C'=0$ $(1)$ in Step 2 he announces the same (opposite) basis that was announced by Alice, however, he always measures his system with the $X$ basis.


\item \textbf{Sifting and announcement:} Alice and Bob announce the bit string in the $X$ basis, which corresponds to the events when $C' = 0$ and $Z_{C} = 1$.

\end{enumerate}

In this virtual protocol, Alice's and Bob's measurements are postponed because it is useful to clearly define the outcomes of each run of the protocol. Note that, in Step 6 Alice employs the $X$ basis for the measurement, and this is required for the estimation of the phase error rate. It can be seen that the classical and quantum information available to Eve is the same in both the actual and virtual protocols, and also the key generated is identical therefore, the virtual protocol can be used to prove the security of the QKD scheme.

%
%
%

\subsection{Simulation of the key rate for the Lo-Preskill's analysis with imperfectly characterised states}

In this section, we evaluate Lo-Preskill's analysis for the particular device model described in Section \ref{sec:simulation}. Since it considers the BB84 protocol \cite{bennett}, Alice sends Bob the three states in Eq. (\ref{eq:alice_states}) and another state in the $X$ basis, which we assume to be
\begin{equation}
\ket{\omega_{1X}}_B = \cos \bigg( \frac{3 \pi}{4} + \frac{3 \delta}{4} \bigg) \ket{0_Z} + \sin \bigg( \frac{3 \pi}{4} + \frac{3 \delta}{4} \bigg) \ket{1_Z}. 
\label{eq:extrastate}
\end{equation}
This state is obtained by following the same device model as that given in Eq. (\ref{eq:model}) for $\varphi_A = 3\pi/2$, i.e., the phase modulation is proportional to the chosen phase value (see Section \ref{sec:description} for more details). 

In order to estimate the phase error rate we use the coin imbalance $\Delta'$, Eq. (\ref{eq:overlap2}) and Eq. (\ref{eq:eph_delta}). To calculate the secret key rate $R$ we use Eq. (\ref{eq:keyrate}), where the yield of single photons in the $Z$ basis is $Y_Z = P_{Z_A} P_{Z_B} \big[ 4 (1 - \frac{\eta}{2}) p_d + \eta\big]$ for overall transmission efficiency $\eta$ and the bit error rate is defined in Eq. (\ref{eq:biterror}). For simplicity, we assume the probabilities $P_{Z_C} \rightarrow 1$, $P_{Z_A} = P_{Z_B} = \frac{1}{2}$ and for fair comparisons, we use the same channel model used to evaluate the generalised loss-tolerant protocol (see Appendix \ref{app:channel}) and the same experimental parameters. As before, we analyse each of the different source imperfections separately to evaluate how they affect the key generation rate. These are: SPFs parametrised by $\delta$, the phase modulation deviation; the non-qubit assumption where $\theta_{j\beta}$ is the mode dependency; and a THA which depends on the intensity of the back-reflected light $\mu$. The results are shown below.

\begin{figure}[H]
	\centering
	\begin{subfigure}[H]{0.45\textwidth}
	\includegraphics[width=7.85cm]{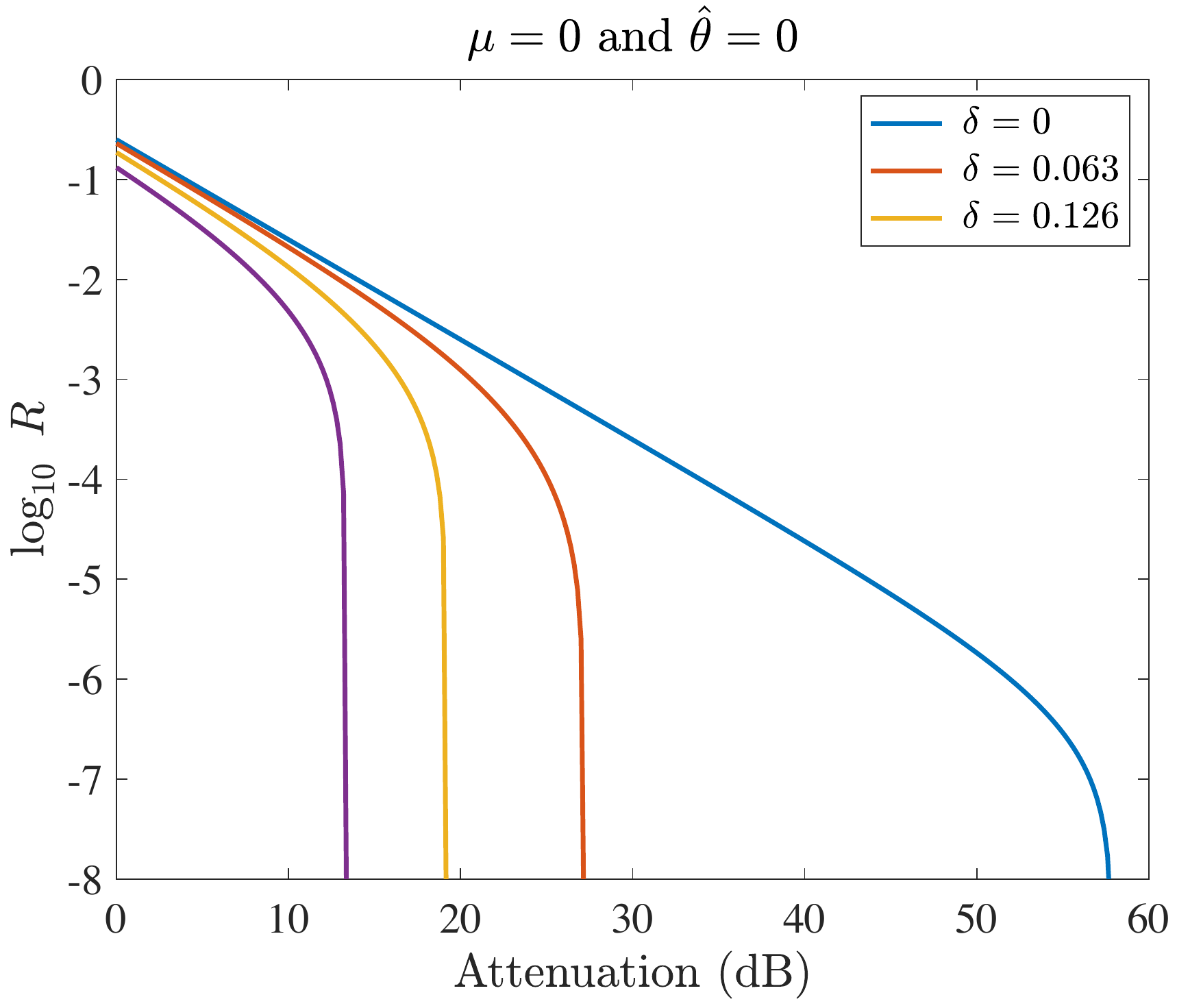} 
	\caption{}
	\end{subfigure}
	\begin{subfigure}[H]{0.45\textwidth}
	\includegraphics[width=7.85cm]{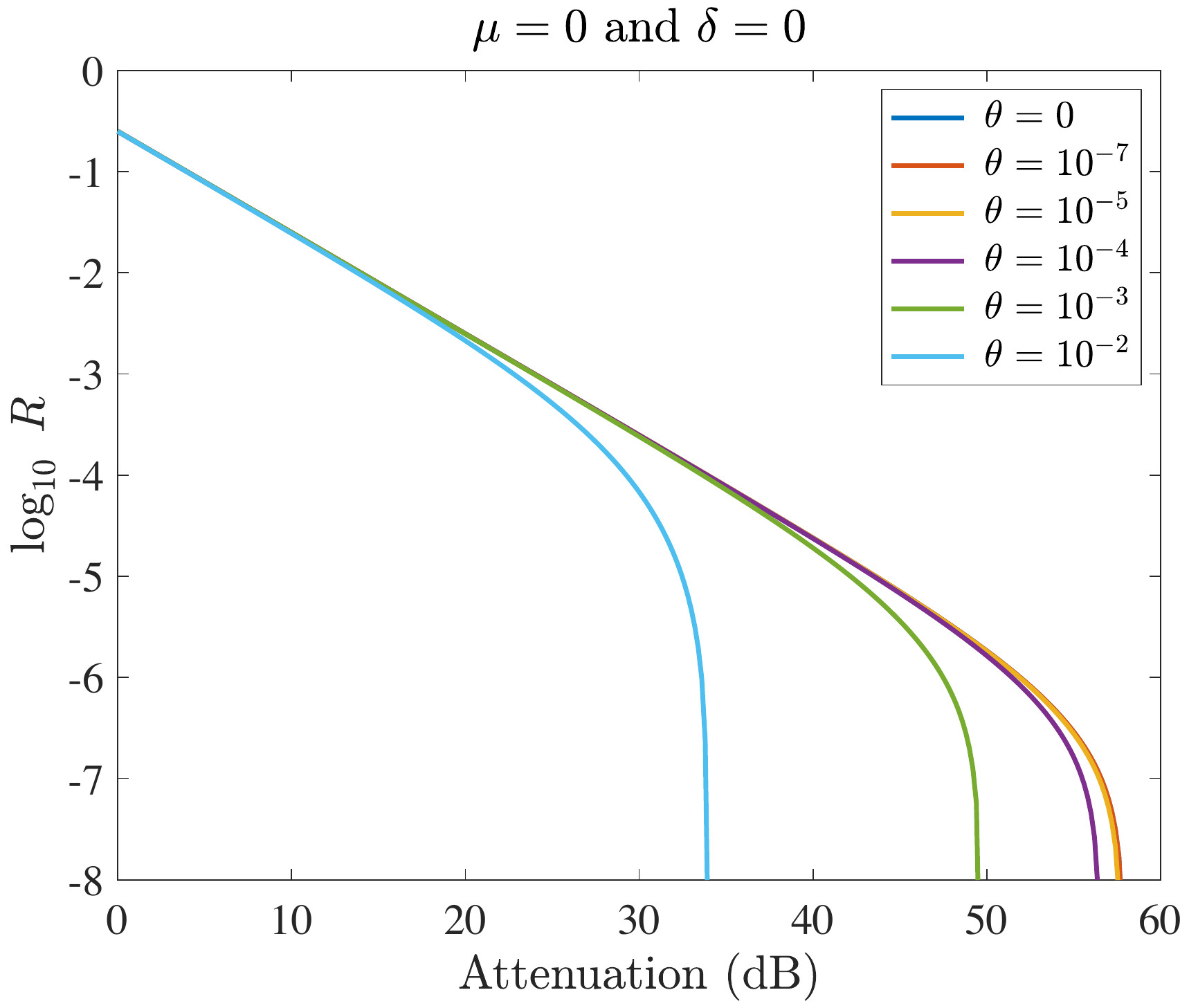}
	\caption{}
	\end{subfigure}
	\begin{subfigure}[H]{0.45\textwidth}
	\includegraphics[width=7.85cm]{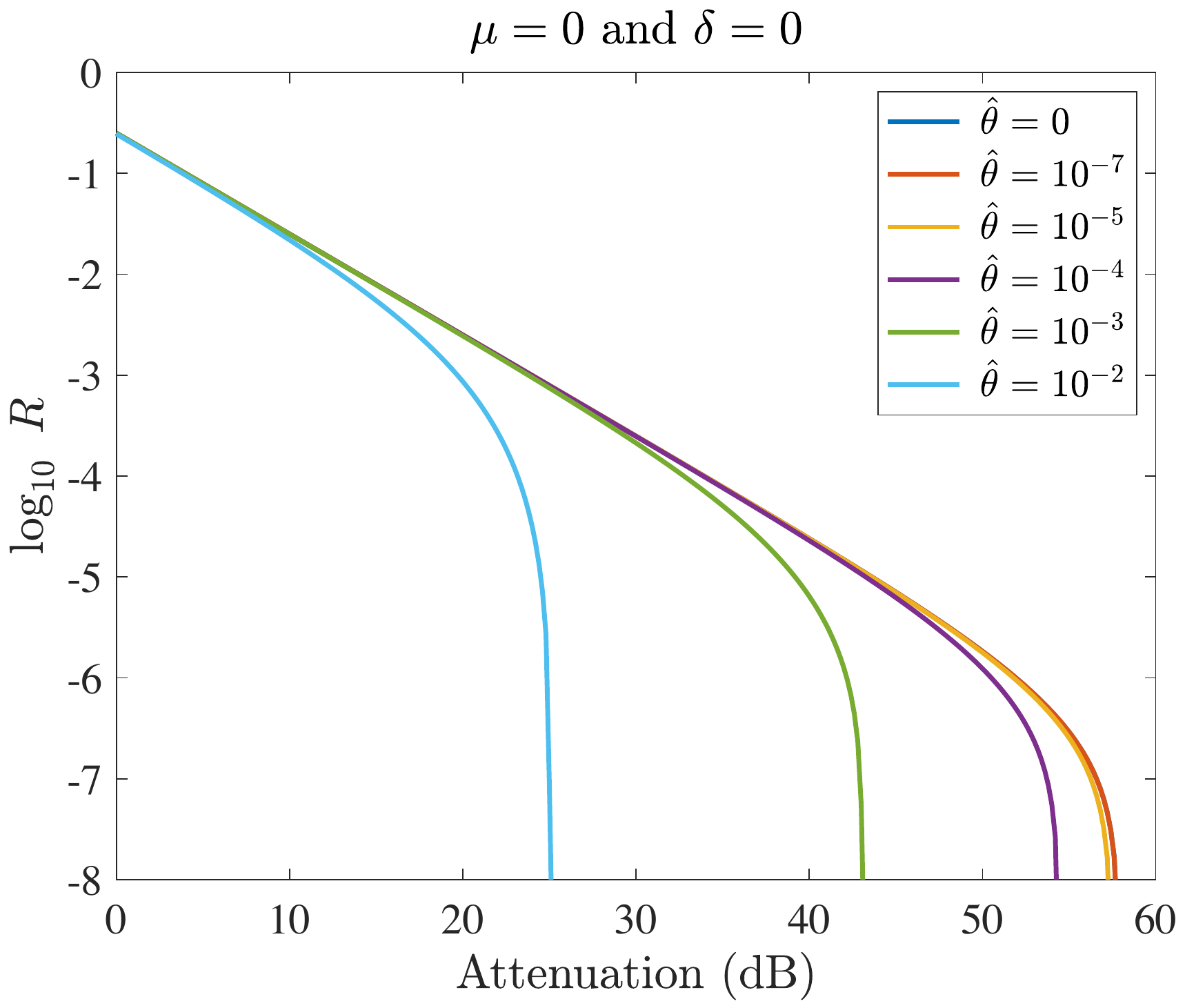}
	\caption{}
	\end{subfigure}
	\begin{subfigure}[H]{0.45\textwidth}
	\includegraphics[width=7.85cm]{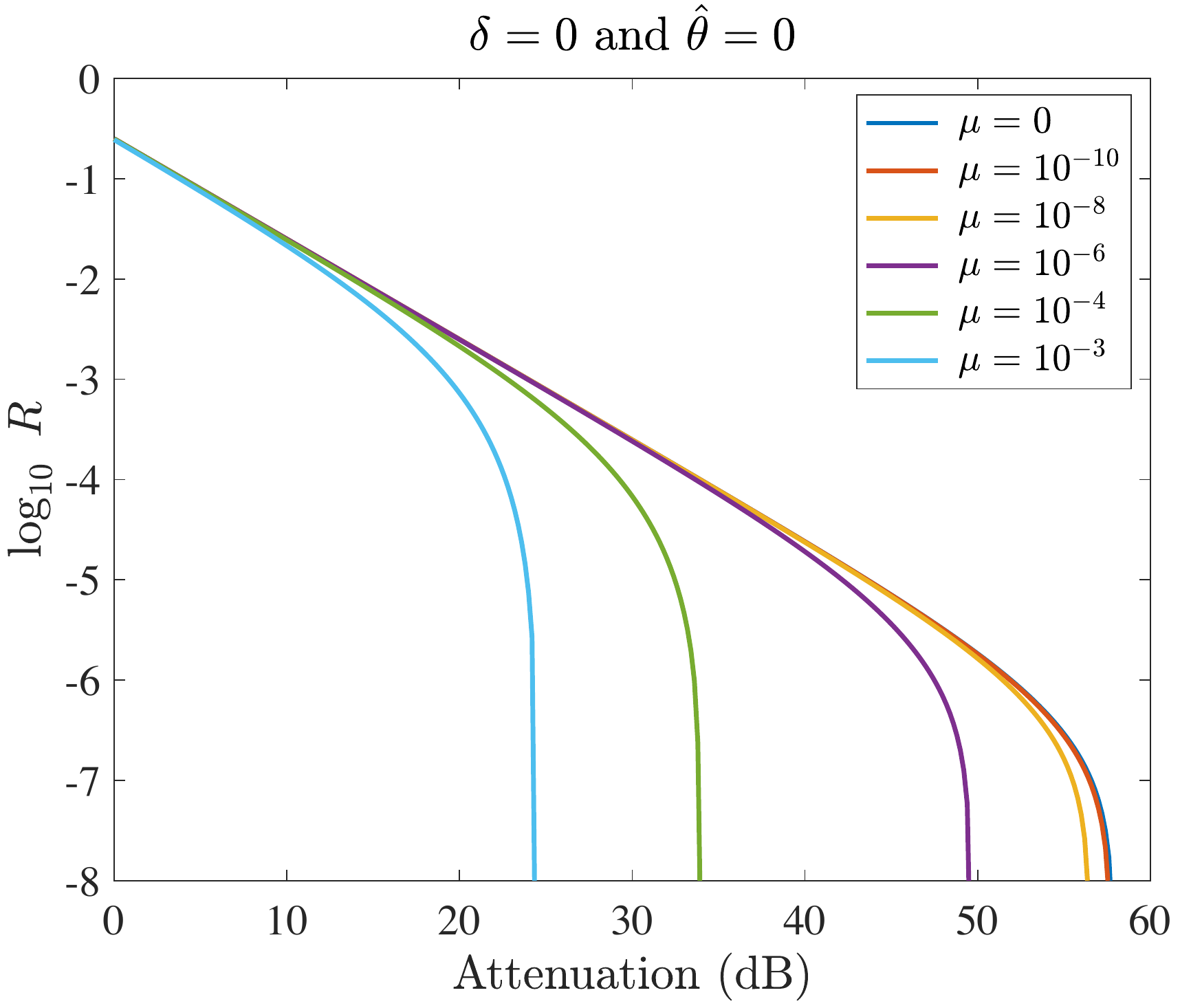}
	\caption{}
	\end{subfigure}
	\caption{Asymptotic secret key rate $R$ against the overall system loss measured in dB, for different values of $\delta$, dependent and independent $\theta$, and $\mu$, using the Lo-Preskill's analysis with imperfectly characterised states. The blue and red curves coincide in some graphs. (a) As the SPF $\delta$ increases, $R$ decreases rapidly. (b) As the setting independent $\theta$ gets larger, the component of vertical polarisation increases and $R$ decreases. (c) When $\theta_{j\beta}$ is setting dependent the key rate $R$ decreases even further. (d) As the intensity of Eve's back-reflected light $\mu$ increases, more information is leaked causing $R$ to decrease.}
	\label{fig:extLP}
\end{figure}

In Fig. \ref{fig:extLP}(a) we can see how $\delta$ greatly affects the secret key rate $R$ as it increases. The explanation for this lies in the fact that Lo-Preskill's method assumes the worst case scenario, namely, the detected signals are chosen to maximise the imbalance of the coin. This assumption is required since not all signals emitted by Alice are detected, thus Eve can exploit these losses to enhance the basis dependence of the detected signals without being exposed. These results contrast with the generalised loss-tolerant protocol, as shown in Fig. \ref{fig:lossTolerant}(a), where $R$ remains almost the same independently of $\delta$.

Fig. \ref{fig:extLP}(b) evaluates the setting independent $\theta$, that is, when $\theta$ is independent of Alice's encoding. As $\theta$ increases the key rate is approximately the same but for values of $\theta \gtrapprox 10^{-4}$ it starts to decrease. However, when we compare it with Fig. \ref{fig:lossTolerant}(b) we see that it decreases slower than in the generalised loss-tolerant protocol. When we consider a setting dependent $\theta_{j\beta}$ instead, the $R$ deteriorates even further, as shown in Fig. \ref{fig:extLP}(c). This result is expected since Eve is now able to better distinguish the states sent by Alice. 

Finally, in Fig. \ref{fig:extLP}(d) we can see how the THA affects the secret key rate. An increase in $\mu$ results in a lower $R$ but, it decreases at a slower rate when compared with the generalised loss-tolerant protocol in Fig. \ref{fig:lossTolerant}(d). As observed, no key can be obtained around $\mu \gtrapprox 10^{-3}$ and the secret key rate is roughly the same for values of $\mu \lessapprox 10^{-8}$. 

From these results we can conclude that the Lo-Preskill's analysis is highly affected by SPF but it is  more resistant to the non-qubit assumption and THA when compared to the generalised loss-tolerant protocol. In order to see this comparison more clearly we refer the reader back to Section \ref{sec:simulation}.

\section{Channel model}
\label{app:channel}
The three states sent by Alice in Eq. (\ref{eq:alice_states}) can be expressed in terms of creation operators as follows

\begin{equation}
\begin{split}
& \ket{\omega_{0Z}}_B = \frac{1}{\sqrt{2}} (\hat{a}^{\dagger}_r + \hat{a}^{\dagger}_s ) \ket{v}, \\
& \ket{\omega_{1Z}}_B = \frac{1}{\sqrt{2}} (\hat{a}^{\dagger}_r - e^{i\delta} \hat{a}^{\dagger}_s ) \ket{v}, \\
& \ket{\omega_{0X}}_B = \frac{1}{\sqrt{2}} (\hat{a}^{\dagger}_r + i e^{i\delta/2} \hat{a}^{\dagger}_s )\ket{v}, \\
\end{split}
\end{equation}

\noindent where $\hat{a}^{\dagger}_r$ and $\hat{a}^{\dagger}_s$ are creation operators for a photon in the reference and signal pulses respectively. Using these states, we can simulate the QKD protocol as shown in Fig. (\ref{fig:protocol}). We assume that the detectors D0 and D1 have the same detection efficiency and that the dark counts $p_d$ are independent of the incoming signals. The overall transmission efficiency of the system can be expressed as $\eta= \eta_{channel} \eta_{detector}$, where $\eta_{channel} = 10^{-\alpha l/10}$, $\alpha$ in dB/Km is the fiber loss coefficient and $l$ is transmission distance in Km. Note that, only half of the pulses interfere in Bob's lab. Furthermore, we do not assume misalignment in the channel. For simplicity, the non-qubit assumption and the THA are not included in this model because we assume that they are not affected by the channel. By neglecting the terms associated with $p_d^2$, the results obtained are

\begin{align*}
& Y_{0X,0Z}^{(Z)}  \approx ~ P_{0Z} P_{X_B} \bigg[\Big(1-\frac{\eta}{2}\Big) p_d + \frac{\eta}{4} \Big(1 + \sin \frac{\delta}{2}\Big) \Big(1 - \frac{p_d}{2}\Big) + \frac{\eta}{8} \Big(1 - \sin \frac{\delta}{2}\Big)p_d \bigg], \\
& Y_{1X,0Z}^{(Z)}  \approx ~ P_{0Z} P_{X_B} \bigg[ \Big(1-\frac{\eta}{2}\Big) p_d + \frac{\eta}{8} \Big(1 + \sin \frac{\delta}{2}\Big) p_d + \frac{\eta}{4} \Big(1 - \sin \frac{\delta}{2} \Big)\Big(1 - \frac{p_d}{2}\Big)\bigg] , \\
& Y_{0Z,0Z}^{(Z)}  \approx ~ P_{0Z} P_{Z_B} \bigg[\Big(1-\frac{\eta}{2}\Big) p_d + \frac{\eta}{4} (1 + \cos \delta) \Big(1 - \frac{p_d}{2}\Big) + \frac{\eta}{8} (1 - \cos \delta)p_d \bigg], \\
& Y_{1Z,0Z}^{(Z)}  \approx ~ P_{0Z} P_{Z_B} \bigg[ \Big(1-\frac{\eta}{2}\Big) p_d + \frac{\eta}{8} (1 + \cos \delta) p_d + \frac{\eta}{4} (1 - \cos \delta) \Big(1 - \frac{p_d}{2}\Big)\bigg], \\
& Y_{0X,1Z}^{(Z)}  \approx ~ P_{1Z} P_{X_B} \bigg[\Big(1-\frac{\eta}{2}\Big) p_d + \frac{\eta}{4} \Big(1 - \sin \frac{3\delta}{2}\Big) \Big(1 - \frac{p_d}{2}\Big) + \frac{\eta}{8} \Big(1 + \sin \frac{3\delta}{2}\Big)p_d \bigg], \stepcounter{equation}\tag{\theequation} \\
& Y_{1X,1Z}^{(Z)}  \approx ~ P_{1Z} P_{X_B} \bigg[ \Big(1-\frac{\eta}{2}\Big) p_d + \frac{\eta}{8} \Big(1 - \sin \frac{3\delta}{2}\Big) p_d + \frac{\eta}{4} \Big(1 + \sin \frac{3\delta}{2}\Big)  \Big(1 - \frac{p_d}{2}\Big)\bigg]  , \\
& Y_{0Z,1Z}^{(Z)}  \approx ~ P_{1Z} P_{Z_B} \bigg[\Big(1-\frac{\eta}{2}\Big) p_d + \frac{\eta}{4} (1 - \cos 2\delta) \Big(1 - \frac{p_d}{2}\Big) + \frac{\eta}{8} (1 + \cos 2\delta)p_d \bigg], \\
& Y_{1Z,1Z}^{(Z)}  \approx ~ P_{1Z} P_{Z_B} \bigg[ \Big(1-\frac{\eta}{2}\Big) p_d + \frac{\eta}{8} (1 - \cos 2\delta) p_d + \frac{\eta}{4} (1 + \cos 2\delta) \Big(1 - \frac{p_d}{2}\Big)\bigg], \\
& Y_{0X,0X}^{(X)}  \approx ~ P_{0X} P_{X_B} \bigg[\Big(1-\frac{\eta}{2}\Big) p_d + \frac{\eta}{4} (1 + \cos \delta) \Big(1 - \frac{p_d}{2}\Big) + \frac{\eta}{8} (1 - \cos \delta)p_d \bigg], \\
& Y_{1X,0X}^{(X)}  \approx ~ P_{0X} P_{X_B} \bigg[ \Big(1-\frac{\eta}{2}\Big) p_d + \frac{\eta}{8} (1 + \cos \delta) p_d + \frac{\eta}{4} (1 - \cos \delta) \Big(1 - \frac{p_d}{2}\Big)\bigg]. \\
\end{align*}

\end{document}